\documentclass{JHEP3}

\usepackage{amsmath}
\usepackage{epsfig,multicol,bbm}
\usepackage{graphicx}
\usepackage{bm}
\usepackage{multirow}
\usepackage{xspace}

\newcommand{\nn}{\nonumber} 
\newcommand{\bn}{{\bar n}}

\newcommand{\be}{\begin{equation}}
\newcommand{\ee}{\end{equation}}

\newcommand{\as}{\alpha_s}

\newcommand{\kt}{\textrm{k}_\textrm{T}}
\newcommand{\akt}{\textrm{ak}_\textrm{T}}

\newcommand{\tL}{\mathsf{L}}

\newcommand{\ca}[1]{\mathcal{#1}}
\newcommand{\ord}[1]{\mathcal{O}(#1)}
\newcommand{\e}{\epsilon}

\newcommand{\wt}{\widetilde}
\newcommand{\cM}{\wt{\ca{M}}}

\newcommand{\io}{\text{IO}}
\newcommand{\oo}{\text{OO}}
\newcommand{\CA}{\text{C/A}}
\newcommand{\alg}{\text{alg}}

\newcommand{\eq}[1]{Eq.~\eqref{eq:#1}}
\newcommand{\eqs}[2]{Eqs.~\eqref{eq:#1} and \eqref{eq:#2}}

\renewcommand{\sec}[1]{Sec.~\ref{sec:#1}}
\newcommand{\ssec}[1]{Sec.~\ref{ssec:#1}}
\newcommand{\fig}[1]{Fig.~\ref{fig:#1}}

\allowdisplaybreaks[3]

\DeclareMathOperator{\Li}{Li}

\title{Abelian Non-Global Logarithms from Soft Gluon Clustering}

 \author{
Randall Kelley \\
Department of Physics, Harvard University, Cambridge, Massachusetts, 02138, USA \\ 
E-mail: \email{rkelley@physics.harvard.edu}
}

\author{
Jonathan R. Walsh and Saba Zuberi \\
Theoretical Physics Group, Ernest Orlando Lawrence Berkeley National Laboratory, \\
and Center for Theoretical Physics, University of California, Berkeley, CA 94720, USA \\
E-mail: \email{jwalsh@lbl.gov}, \email{szuberi@lbl.gov}
 }

\abstract{
Most recombination-style jet algorithms cluster soft gluons in a complex way.  This leads to correlations in the soft gluon phase space and introduces logarithmic corrections to jet cross sections.  The leading Abelian clustering logarithms occur at least at next-to leading logarithm (NLL) in the exponent of the distribution, and we show that new clustering effects contributing at NLL likely arise at each order.  Therefore we find that it is unlikely that clustering logs can be resummed to NLL. Clustering logarithms make the anti-$\kt$ algorithm theoretically preferred, for which they are power suppressed. They can arise in Abelian and non-Abelian terms, and we calculate the Abelian clustering logarithms at $\ca{O} (\as^2)$ for the jet mass distribution using the Cambridge/Aachen and $\kt$ algorithms, including jet radius dependence, which extends previous results.  We find that previously identified logarithms from clustering effects can be naturally thought of as a class of non-global logarithms (NGLs), which have traditionally been tied to non-Abelian correlations in soft gluon emission. 
}


\begin{document}


\section{Introduction}

Events with jets are a key signature at collider experiments.  Jet cross sections are used to measure jet rates and the structure of jets, providing leverage to discriminate between different signals.  Accurate theoretical predictions for jet cross sections require understanding, among other things, the perturbative structure of these rates.  This is often complicated by clustering effects of the jet algorithm, the presence of many scales in the measurement, and the need to resum large logarithms.

The perturbative series for a jet cross section is governed by two things: the structure of QCD matrix elements and the specific observable that is measured.  Measurements that probe the structure of jets can generate complex perturbative series that depend on many kinematic scales.  Even basic jet observables depend on the choice of jet algorithm and associated parameters, such as the jet radius $R$ and the veto $\Lambda$ on soft jets.

When a measurement is sensitive to soft and collinear radiation, large logarithms can appear in the perturbative series and spoil a naive expansion in $\as$.  This occurs for a wide range of jet observables of interest. In this paper we focus on the measurement of $\rho$ and $\Lambda$ in $e^+e^- \to 2$ jet events, where $\rho$ is the sum of the two jet invariant masses scaled by the center of mass energy $Q$, $\rho \equiv (m_1^2 + m_2^2)/Q^2$, and $\Lambda$ is the total energy outside of the two jets.  The cross section $d^2\sigma/ d \rho d\Lambda$ is sensitive to soft and collinear radiation when $\rho \ll Q$ and has leading logarithms of the form $\as^n \ln^{2n} \rho$.  These logs must be resummed to regain perturbative accuracy, and this is achieved through factorization and renormalization group evolution (RGE) both in perturbative QCD \cite{Sterman:1995fz,Contopanagos:1996nh} and in soft-collinear effective theory (SCET) \cite{Bauer:2000ew,Bauer:2000yr,Bauer:2001ct,Bauer:2001yt,Bauer:2002nz}.  Factorization separates the perturbative cross section into separately calculable pieces that depend on different energy scales.  For the total dijet mass distribution the hierarchy of scales when $\rho \ll Q$ separates the physics of the event into distinct processes, given schematically by
\be\label{eq:factschem}
\frac{d^2\sigma}{d \rho d \Lambda} = \sigma_0 \, H(Q) \,J_1 (Q\,\sqrt{\rho} ) \otimes J_2 (Q\,\sqrt{\rho}) \otimes S( Q \rho, \Lambda)
\ee
The short distance interaction occurs at a scale of order $Q$ in the hard function $H$ and produces the high energy partons that give rise to the jets.  This is far above the jet mass $Q \sqrt{\rho}$ which is the scale associated with the collinear evolution in the jet function $J$.  Soft radiation inside of and between jets exists at an even lower energy, $Q \rho$, described by the soft function, $S$.

Non-global logarithms (NGLs) arise when the measurement groups the soft radiation in distinct regions of phase space, each of which contributes to a different observable \cite{Dasgupta:2001sh,Dasgupta:2002bw}.   In the total dijet mass cross section, there are NGLs of the scale ratio of the dijet mass $\rho$ and the veto $\Lambda$ on additional jets.  These start at order $\as^2 \ln^2 Q\rho / \Lambda$ and continue to contribute at next-to-leading log\footnote{We will use a standard log counting scheme, counting in the exponent of the distribution.  In this case $\text{N}^k\text{LL}$ terms are of order $\as^n \ln^{n+1-k}$ for all $n$ (with $k=0$ for leading log).  We will also refer to log counting in the distribution itself, which sums fewer logs at an equivalent order.} (NLL) at higher orders.  NGLs exist due to correlations in non-Abelian matrix elements for soft gluon emission. In factorization theorems of the form of \eq{factschem}, they enter as terms in the soft function not associated with the anomalous dimension.  Because they are connected with scales not explicitly separated by the functions in the factorization theorem, they are not resummed by renormalization group evolution (RGE).

\begin{figure}[t]{
	\begin{center}
	\includegraphics[width=.7\textwidth]{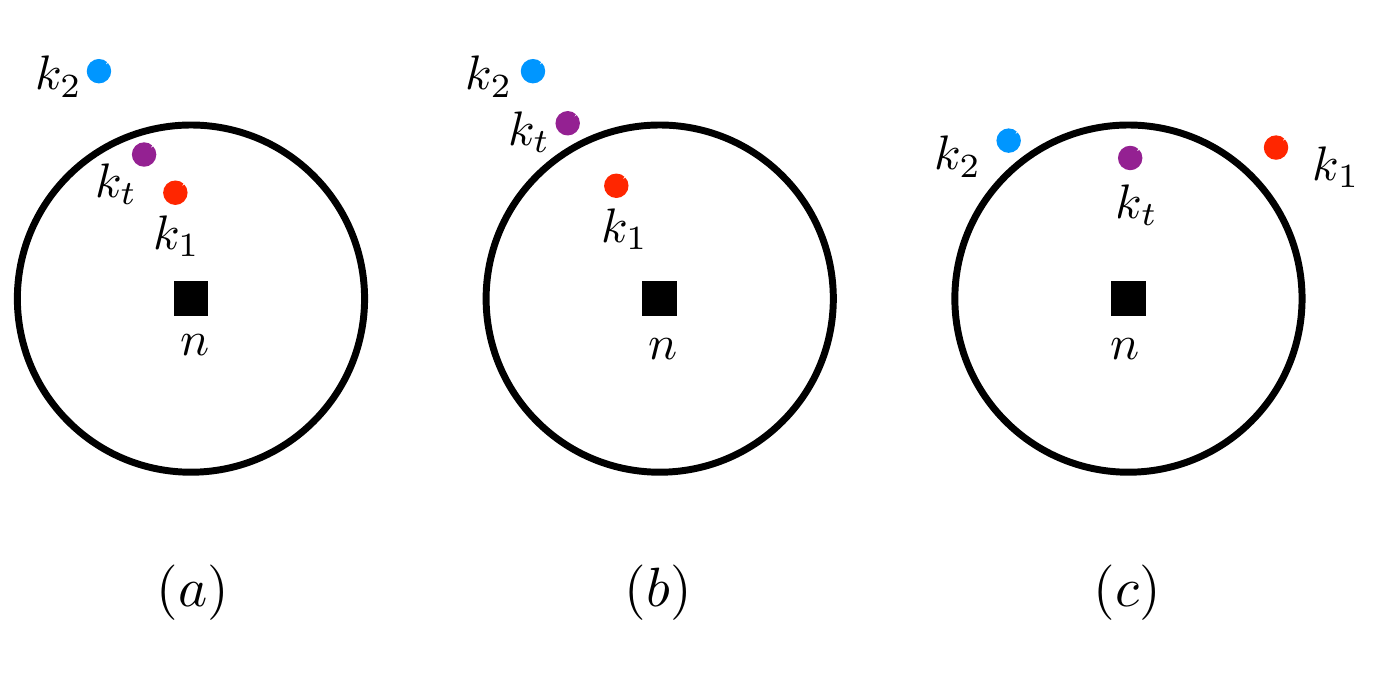} 
	  \end{center} 
	  \vspace{-2em}
{ \caption[1]{The two particle configurations that change the jet boundary. Particles $k_1$ and $k_2$ are combined across the jet boundary to give the combined pair $k_t=k_1+k_2$.  Figures (a) and (b) show the in-out contribution where particles are clustered across the jet boundary and (c) shows the out-out contribution where two particles out of the jet are combined into the jet.}
  \label{fig:CAps}} }
\end{figure}

Recombination-style jet algorithms are a common choice to define jets.  When performing theoretical calculations using these algorithms, the phase space constraints quickly become very complex as the number of final state partons increases. When a pair of particles is clustered, one or both of these partons may be pulled into (or out of) the jet, changing the jet boundary and producing correlations between particles in the measurement of the final state.  The schematic configurations for 2 particles are shown in \fig{CAps}(a)-(c).
For observables sensitive to soft radiation, clustering effects can lead to logarithms in the cross section.  These are termed \textit{clustering logarithms} and have been calculated for specific jet cross sections \cite{Banfi:2005gj,Delenda:2006nf,KhelifaKerfa:2011zu}.

\begin{figure}[t]{
	\begin{center}
	\includegraphics[width=\textwidth]{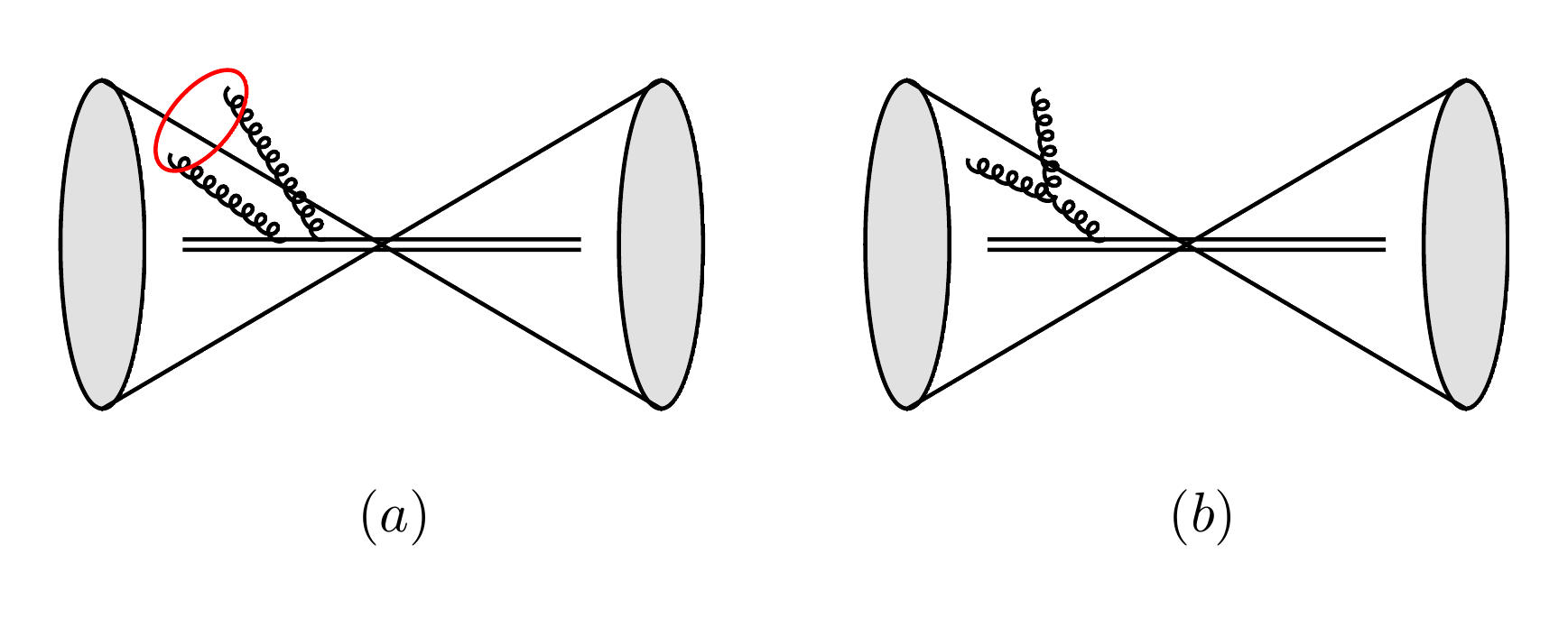} 
	  \end{center} 
	  \vspace{-2em}
{ \caption[1]{Clustering logs and NGLs arise from correlations between soft gluons in separate regions of phase space that are sensitive to different physical scales.  In (a), clustering logs arise from correlations in the measurement due to clustering.  In (b), NGLs arise from correlations in the non-Abelian matrix element for multiple soft gluon emission.  The similarity in origin leads us to call both types of logs NGLs.}
  \label{fig:NGLdiagram}} }
\end{figure}

In this paper we will show that clustering logarithms arise at least at NLL and that they give rise to a new and unique contribution at each order in $\as$. It is therefore unlikely that clustering logarithms can be resummed to NLL, which limits the perturbative accuracy of jet observables involving algorithms such as Cambridge/Aachen (C/A) or $\kt$ \cite{Catani:1991hj,Catani:1993hr,Ellis:1993tq,Dokshitzer:1997in}, for which clustering effects arise. This impacts a wide range of processes involving jets both at $e^+ e^- $ and hadron colliders and makes the anti-$\kt$ algorithm (\cite{Cacciari:2008gp}) and event shape observables such as $N$-jettiness (\cite{Stewart:2010tn}) theoretically preferred, for which clustering logarithms do not arise. We will also explore the connection between clustering logarithms and NGLs in SCET. Both arise in the soft function, and come from correlations between soft gluons.  The schematic form of the soft function is
\be
S = \int d\Phi \, \ca{A}(\Phi) \, \ca{M}(\Phi) \,.
\ee
NGLs come from correlations in the squared matrix element, $\ca{A}$, which occur only in non-Abelian terms.  Clustering logs come from correlations in the measurement, $\ca{M}$, and contain terms that violate Abelian exponentiation.  In \fig{NGLdiagram}, typical configurations that give rise to clustering and non-global logs are shown for a dijet event at $\ca{O}(\as^2)$.
 The two types of logarithms share several characteristics:
\begin{itemize}
\item They arise from correlations between soft gluons, in the matrix element for NGLs and in the measurement for clustering logs.
\item They are present when the measurement is divided into regions where different physical scales constrain the phase space.  When correlated soft gluons enter into different regions, they produce logs whose argument is the ratio of these scales.
\item They are associated with soft divergences in the matrix element.
\item They are independent of the UV divergences that give rise to the soft function anomalous dimension and are therefore not summed by standard RGE.
\end{itemize}
We find it natural to consider clustering logs as a class of NGLs, and we will call them \textit{clustering NGLs}.

We study clustering NGLs in the total dijet mass distribution for $e^+e^- \to 2$ jets 
\be\label{eq:doublydiff}
\frac{d^2 \sigma}{d\rho \, d\Lambda} \,,
\ee
This observable was defined in \cite{Kelley:2011tj} and called jet thrust, $\tau_\omega$, where $\omega$ was used instead of $\Lambda$. This distribution depends on the jet algorithm, and in \sec{dijet} we give the factorization theorem and discuss the impact of different jet algorithms on clustering logs.  We will show that the anti-$\kt$ algorithm obeys Abelian exponentiation and does not generate clustering NGLs, but that other iterative jet algorithms do.  In \sec{AbelianNGLs} we calculate the $\ca{O}(\as^2)$ Abelian clustering NGLs for the C/A and $\kt$ algorithms away from the small $R$ limit, finding excellent agreement with the Monte Carlo program EVENT2 \cite{Catani:1996jh,Catani:1996vz}. This extends the work of \cite{KhelifaKerfa:2011zu}, which calculated the leading clustering logarithms at $\ca{O}(\as^2)$ in the small $R$ limit for the C/A algorithm.  

In \sec{NGL} we discuss the properties of clustering NGLs and their connection to traditional NGLs, and in \sec{resum} we discuss the structure of Abelian clustering NGLs at higher orders.  Since the Abelian matrix element is simple at any order, we need only understand how the measurement function for a jet algorithm handles an $n$ particle final state. We show that at $\ca{O}(\as^n)$ Abelian clustering NGLs contribute at NLL order, $\as^n \ln^n(Q\rho/\Lambda)$.  Since this is the first order at which $n$ particle clustering effects arise, we expect there will be a new contribution to the coefficient of NLL clustering logarithms at each order in $\as$ that is unrelated to lower order coefficients. We see this explicitly from our results at $\ca{O}(\as^2)$. Therefore, it seems that clustering NGLs cannot be resummed, since determining the cross section at NLL would require the calculation of an infinite number of distinct coefficients. In \cite{KWZinprepMeas} we prove that this is in fact the case by using a novel framework to express the all-orders form of the measurement function. However, NLL resummation using weaker log counting in the distribution itself is still valid, as this counting reorganizes the terms to make the lowest orders (in $\as$) of clustering NGLs the most important. In \sec{conclusions} we present our conclusions.

\section{The Dijet Mass Distribution}
\label{sec:dijet}

The process $e^+ e^- \to 2$ jets provides a simple system to study the perturbative structure of jet cross sections. We focus on the total dijet mass distribution \eq{doublydiff}, which is obtained as follows. For each event, we cluster the final state into jets using a jet algorithm and label the invariant mass of the two most energetic jets $m_1$ and $m_2$.  We measure $\rho = (m_1^2 + m_2^2)/Q^2$ and the total energy $\Lambda$ of particles outside the two leading jets. Additional jets are vetoed by requiring $\Lambda \ll Q$.

Adding the jets' masses together reduces the complexity of the calculations by removing one scale from the problem.  The doubly differential cross section, \eq{doublydiff}, and related observables have been used to study various properties of jet cross sections, including the cross section dependence on the jet algorithm, the jet radius $R$, the ability to simultaneously resum logarithms of ratios of the hard scale $Q$, the jet scale $Q\sqrt{\rho}$, and the soft scales $Q\rho$ and $\Lambda$, and the impact of non-global logarithms \cite{Appleby:2002ke,Appleby:2003sj,Banfi:2005gj,Delenda:2006nf,Ellis:2009wj,Ellis:2010rwa,Cheung:2009sg,Banfi:2010pa,Kelley:2011ng,Hornig:2011tg,Kelley:2011tj,Kelley:2011aa}.

\begin{figure}[t]{
	\begin{center}
	\includegraphics[width=.4\textwidth]{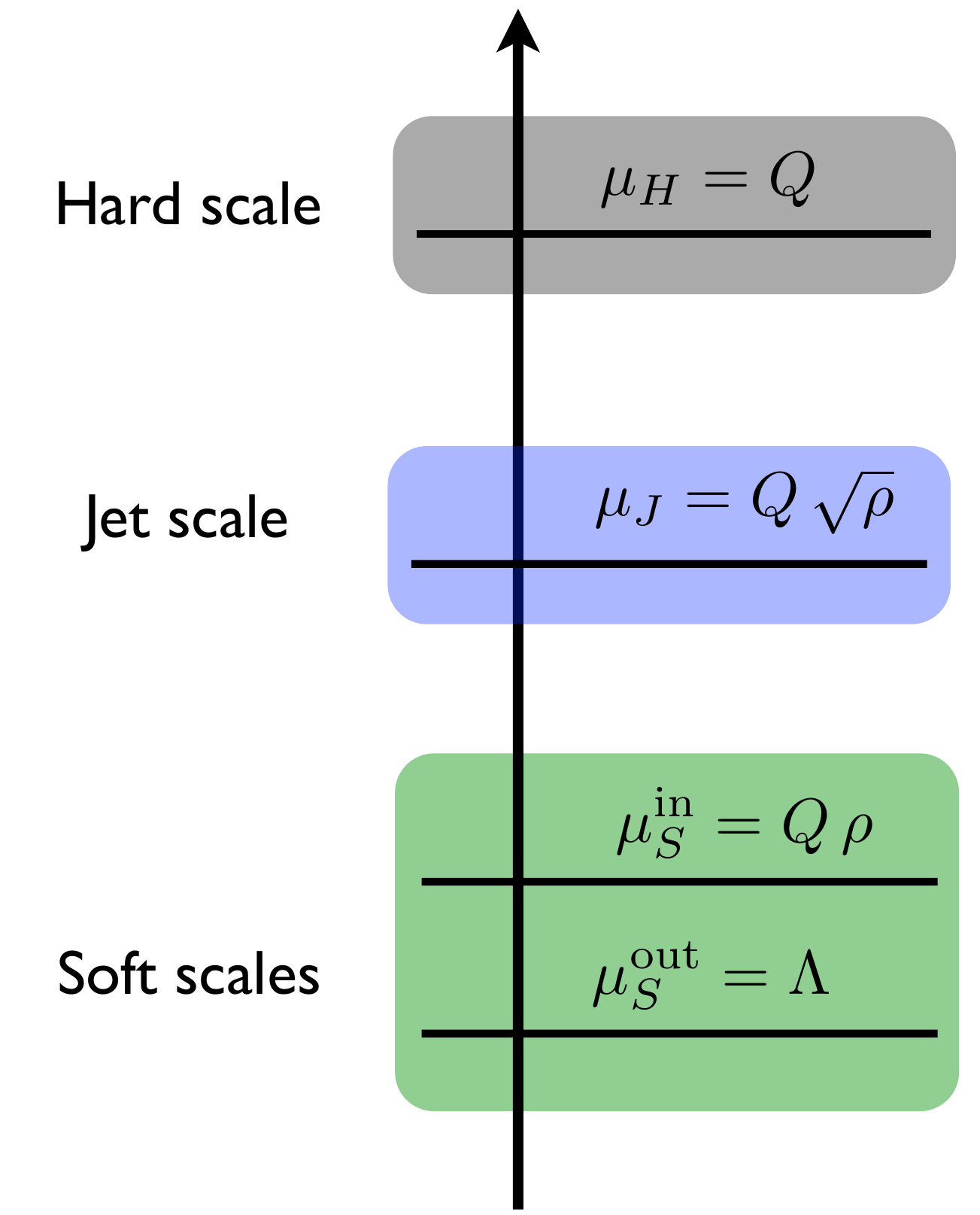}
	\end{center} 
	  \vspace{-2em}
{ \caption[1]{Relevant scales in the cross-section with a measurement on the sum of jet masses with a veto on the radiation outside the jet. The soft function depends on both the scales $\mu_S^{\rm in} \sim Q\, \rho$ and $\mu_S^{\rm out}\sim\Lambda$. }
  \label{fig:scales} } }
\end{figure}

We will study this cross section using the framework of SCET, as well as the double cumulant when convenient,
\be \label{eq:cumulantdef}
\Sigma(\rho_c,\Lambda_c) = \int_0^{\rho_c} d\rho \int_0^{\Lambda_c} d\Lambda \, \frac{d^2 \sigma}{d\rho\, d\Lambda} \,.
\ee
We will work in the dijet limit of small jet mass and small out-of-jet energy,
\be \label{eq:dijetlimit}
\{Q\rho,\Lambda\} \ll Q \sqrt{\rho} \ll Q \,,
\ee
where $Q \sim E_J$ and the final state is dominated by soft and collinear radiation.  SCET can be applied in this regime, and the cross section factorizes into hard, jet, and soft functions \cite{Schwartz:2007ib,Fleming:2007xt,Bauer:2008dt}:
\be \label{eq:sigmafact}
\frac{d^2\sigma}{d\rho \, d\Lambda} = \sigma_0 \, H_2 (Q) \, \int d\rho_n \, d\rho_{\bn} \, d\rho_s \, \delta( \rho - \rho_n - \rho_{\bn} - \rho_s) \, J_n (\rho_n) \, J_{\bn} (\rho_{\bn}) \, S(\rho_s, \Lambda) \,,
\ee
where the labels $n$ and $\bn$ refer to the back-to-back jet directions\footnote{Any four-vector can be decomposed as
\begin{align}
p^\mu =  n\cdot p \frac{ \bn^\mu}{2}+ \bn \cdot p \frac{n^\mu}{2}+p_\perp^\mu \, . \nn
\end{align} 
} $n^\mu,\bn^\mu = (1,\pm \hat{\bf n})$. Each of these functions describe the physics at different scales, depicted in \fig{scales}. 
The hard function describes the short distance interaction at the hard scale $Q$ and is independent of the observable.  The jet function describes collinear evolution of massless partons into jets with mass $m \sim Q\,\sqrt{\rho}$, and the soft function describes global soft radiation in and between jets.  The radiation described by the jet function is collimated along the jet axis.  As long as this collinear radiation is sufficiently narrow compared to the jet radius, satisfying $\rho \ll R^2$, it will be insensitive to the jet boundary and therefore different jet algorithms will cluster these particles equivalently \cite{Ellis:2009wj,Ellis:2010rwa,Jouttenus:2009ns}. In this limit the jet functions $J_{n,\bn}$ are universal up to power corrections for a wide range of processes.  Soft radiation, however, probes the entire final state, and is sensitive to scales in the jet (the scale $Q\rho$) and out of the jet (the scale $\Lambda$, as well as the parameter $R$). It is therefore sensitive to the jet boundary and the soft phase space constraints will differ between jet algorithms, leading us to focus on the soft function.

The soft function is a forward scattering matrix element of soft Wilson lines $Y_{n,\bn}$,
\be\label{eq:Sfun}
S(\rho, \Lambda) = \frac{1}{N_c} \text{Tr} \,\langle 0 \lvert Y_{\bn} \, Y_n^{\dagger} \, \widehat{\ca{M}}(\rho, \Lambda)\, Y_n \, Y_{\bn}^{\dagger} \rvert 0 \rangle \,,
\ee
where the measurement operator $\widehat{\ca{M}}(\rho,\Lambda)$ implements the measurement on the final state, including the jet algorithm.  The measurement acts on the final state of soft particles with momenta $\{ k_s \}$ as follows
\be
\widehat{\ca{M}}(\rho,\Lambda) \, | X_s \rangle = \ca{M} \big(\rho,\Lambda; \{k_s\}\big) \, | X_s \rangle \,,
\ee
where
\be\label{eq:Mpos}
\ca{M} \big(\rho,\Lambda; \{k_s\}\big) = \delta \left(\rho - \frac{1}{Q}\sum_{i\in n{\text{ jet}}} n\cdot k_i - \frac{1}{Q}\sum_{i\in \bn{\text{ jet}}} \bn\cdot k_i\right) \, \delta \left( \Lambda - \sum_{i \,\not\in n,\bn{\text{ jets}}} k_i^0 \right) \,.
\ee
The jet algorithm determines which particles are in the jets, meaning the sums over particles in and out of the jets have phase space constraints imposed by the jet algorithm.

Soft radiation will affect the value of the jet mass, and thus the observable $\rho$; hence, the soft function is convolved with the jet function in the factorization theorem of \eq{sigmafact}.  It will be convenient to express the cross section in Fourier space for both $\rho$ and $\Lambda$, which removes the convolutions and makes the measurement more straightforward:
\begin{align} \label{eq:sigmapos}
\frac{d^2\wt{\sigma}}{dx\, dy} &= \int_{-\infty}^{\infty} d\rho \, d\Lambda \, \frac{d^2\sigma}{d\rho \, d\Lambda} \, \exp( -ix\rho) \exp(- iy\Lambda/Q) \nn \\
&= \sigma_0 \, H_2 (Q) \, \wt{J}_n (x) \, \wt{J}_{\bn} (x) \, \wt{S} (x,y) \,.
\end{align}
The Fourier space jet and soft functions, $\wt{J}$ and $\wt{S}$, are the Fourier transforms of the momentum space functions.  The Fourier space form of the measurement in the soft sector is
\be \label{eq:MeasPos}
\cM \big(x,y; \{k_s\}\big) = \exp\left( -i x\sum_{i\in n{\text{ jet}} } \frac{n\cdot k_i}{Q} \right) \exp\left( -i x\sum_{i\in \bn{\text{ jet}} } \frac{\bn\cdot k_i}{Q} \right) \exp\left( -i y\sum_{i \,\not\in n,\bn{\text{ jets}} } \frac{k_i^0}{Q} \right) \,.
\ee
This form is useful as it removes the convolution in \eq{Mpos} between different particles' momenta, and will be helpful in understanding NGLs.

\subsection{Jet Algorithms and Boundary Clustering}
\label{ssec:jetalg}

Jet algorithms are used to identify energetic clusters of radiation.  Inclusive recombination algorithms, which we study here, build the jet through pairwise clustering of particles.  These algorithms utilize two kinds of metrics; a pairwise metric $\rho_{ij}$ between two particles and a single particle metric $\rho_i$ for an individual particle.  The algorithm works recursively, finding the minimum of all pairwise and single particle metrics.  If the minimum is a pairwise metric, then that pair is merged by adding their four momenta, $p_{i+j} = p_i + p_j$.  If the minimum is a single particle metric, then that particle is promoted to a candidate jet.  This is repeated until all particles have been clustered into candidate jets.  Finally, an energy or $p_T$ cut is used to veto soft jets and select final state jets.

The $\kt$ class is a common set of jet algorithms \cite{Catani:1991hj,Catani:1993hr,Ellis:1993tq,Dokshitzer:1997in,Cacciari:2008gp}.  The metric for these algorithms is parameterized by a number $\alpha$.  For $e^+e^-$ collisions, the metrics take the form:
\begin{align}
\rho_{ij} &= 2\min\left( E_i^{2\alpha} , E_j^{2\alpha} \right) (1-\cos\theta_{ij}) \,, \nn \\
\rho_i &= 2 \, E_i^{2\alpha} (1-\cos R) \,,
\end{align}
where $\theta_{ij}$ is the angle between particles $i$ and $j$.  The radius $R$ sets the size of the jet, and is the maximum angle between two particles for which a single clustering can occur. This class includes the $\kt$, Cambridge/Aachen (C/A), and anti-$\kt$ algorithms:
\begin{align}\label{eq:algs}
\alpha = 1 : \kt \,, \qquad
\alpha = 0 : \text{C/A} \,,  \qquad
\alpha = -1 : \text{anti-$\kt$} \,.
\end{align}
For a single soft particle, these algorithms give the same phase space constraints; the jet boundary is simply a cone of radius $R$ around the jet direction.  The position space measurement function for a single particle for $\kt$, C/A and anti-$\kt$ is 
\be
\cM^{(1)} (x,y ; k) 
= \exp\left( 
	-i x\,  \frac{n\cdot k}{Q} \, \Theta_n^{k}
	-i x\, \frac{\bn \cdot k}{Q} \, \Theta_{\bn}^{k}
	-i y\, \frac{k^0}{Q} \, \Theta_{\rm out}^{k}
	\right)  \,,
\ee 
where
\begin{align}
k \in n{\text{ jet}}       &: \Theta^{k}_{n}      = \theta( R - \theta_{kn} ) \,, \nn \\
k \in \bn{\text{ jet}} &: \Theta^{k}_{\bn}  = \theta( \theta_{kn} - (\pi - R ) ) \,, \nn \\
k \notin n,\bn{\text{ jets}} &: \Theta^{k}_{\rm out}  = \theta( (\pi - R) - \theta_{kn}  ) \, \theta( \theta_{kn} - R ) \,, 
\end{align}

In the small $\rho$ and small $\Lambda$ regime, when $\rho \ll R^2$, collinear radiation in the jet is parametrically narrower than the jet size and is insensitive to clustering effects of the algorithm. The jet boundary is therefore determined by soft radiation.  The behavior of the $\kt$ class of algorithms have been previously analyzed in the context of SCET, where the metrics determine the characteristic sequence of clustering and the effect on the jet boundary \cite{Walsh:2011fz}.  Soft partons will either first cluster with collinear radiation or  among themselves, depending on the algorithm.  In the former case, the phase space restrictions take the form of a cone of radius $R$ around the jet axis.  When soft particles merge among themselves, they can change the geometry of the jet boundary away from a cone.  
We refer to this as
\begin{align}
\textit{boundary clustering :  }& \text{clustering of soft particles across the jet boundary of radius $R$}  \nn \\
&\text{around the jet direction.} \nn
\end{align}
While clustering purely inside or outside of the jet of course still occurs, it does not change the measurement function, \eq{MeasPos}, and therefore we do not need to consider it.  Boundary clustering however is relevant because it leads to a change in the observable. As an example consider the configuration with two nearby soft gluons shown in \fig{CAps}(a) and (b), where gluon 1 is inside the cone of the $n$ jet ($\theta_{1n} < R$) and gluon 2 is outside this cone ($\theta_{2n} > R$).  If these gluons are not clustered, then gluon 1 is in the jet and contributes to the $\rho$ measurement and gluon 2 is out of the jet and contributes to the $\Lambda$ measurement.  However, if the gluons are clustered, then both gluons will either be in the jet, contributing to $\rho$, or be out of the jet, contributing to $\Lambda$, as shown in \fig{CAps}(a) and  \fig{CAps}(b), respectively.  Thus, boundary clustering changes the observable and can introduce logarithmic corrections in the soft regime.

\begin{figure}[t]{
	\begin{center}
	\includegraphics[width=.7\textwidth]{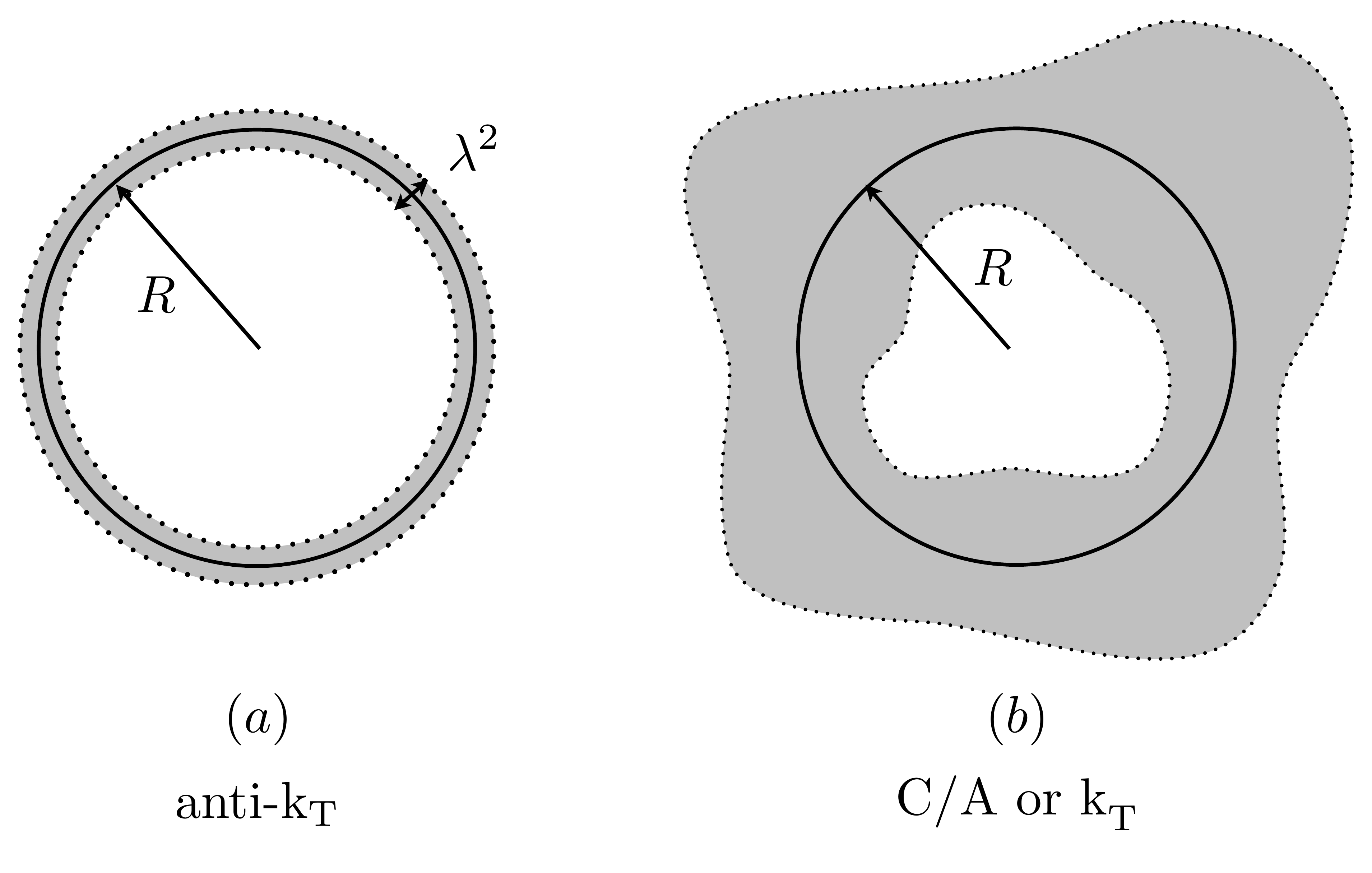} 
	  \end{center} 
	  \vspace{0em}
{ \caption[1]{ The region of soft particle boundary clustering is shown schematically in grey for (a) anti-$\kt$ and (b) C/A, $\kt$. This corresponds to the correction to a cone of radius $R$ in the measurement function given by $\Delta \cM_\alg$ for each algorithm. For anti-$\kt$ this region of phase space is power suppressed by a factor $\lambda^2 \sim \rho \ll 1$. For C/A and $\kt$ the region of soft particle boundary clustering corresponds to an $\ca{O}(1)$ region of phase space.}
  \label{fig:BoundClusReg}} }
\end{figure}

Boundary clustering gives correlations in the phase space constraints for soft particles.  It is useful to express the measurement function in \eq{MeasPos} for $n$ particles in terms of a product of independent single particle constraints plus a correction $\Delta \cM_\alg$ due to boundary clustering:
\be\label{eq:MeasN}
\cM^{(n)}_\alg (x,y,\{ k_1,\ldots,k_n\}) 
= \prod_{i=1}^n \cM^{(1)} (x,y,k_i) 
+ \Delta \cM_\alg (x,y,\{ k_1,\ldots,k_n\}) \,.
\ee
In \fig{BoundClusReg}, we show the schematic size of the region contributing to $\Delta \cM_\alg$ for the anti-$\kt$, C/A, and $\kt$ algorithms.
Since the anti-$\kt$ metric is weighted by the inverse energy, in the dijet limit the algorithm first clusters particles with scaling collinear to the jet directions.  Then clustering between soft and collinear radiation typically occurs before clustering among soft particles, meaning that the jet boundary for soft particles is only set by the jet axis and is a cone of radius $R$ up to power corrections \cite{Walsh:2011fz}. As a result, for the anti-$\kt$ algorithm, the region of phase space where boundary clustering can take place ($\Delta \cM_{\akt}$) is power suppressed and the measurement function in \eq{MeasN} factorizes:
\begin{align}
\label{eq:MeasPosAT}
\cM^{(n)}_{\akt} \big(x,y; \{k_s\}\big) 
&= \prod_{i=1}^n \cM^{(1)} (x,y ; k_i)\,. 
\end{align}
The measurement function also factorizes in this way for global event shape observables such as thrust, angularities, or $N$-jettiness \cite{Schwartz:2007ib,Bauer:2008dt,Hornig:2009vb,Stewart:2010tn}.  For the C/A and $\kt$ algorithms, boundary clustering can change the jet boundary by an $\ca{O}(1)$ amount and therefore contributes at leading power.  Since the C/A metric depends only on angle, pairs of soft particles that are closer to each other than the jet axis will cluster.  For the $\kt$ algorithm, this clustering is enhanced, since the metric is weighted to preferentially merge softer particles earlier in the algorithm.

The correction term $\Delta \cM_{\alg}$ in the measurement function leads to a correction in the soft function,
\be\label{eq:wtSdiff}
\wt{S}_\alg(x,y) = \wt{S}_{\akt}(x,y) +\Delta \wt{S}_\alg(x,y) \,,
\ee
and correspondingly in the cross section,
\be
\frac{d^2\wt{\sigma}_\alg}{dx\, dy} = \frac{d^2\wt{\sigma}_{\akt} }{dx\, dy} + \frac{d^2 \Delta \wt{\sigma}_\alg}{dx\, dy} \,.
\ee
Boundary clustering will generate clustering logs in the cross section.  We will focus on the Abelian terms, where the analysis is simplest.

\subsection{Abelian Exponentiation}
\label{ssec:abelexp}

At $\ca{O}(\as^n)$, a general Fourier space soft function contribution has the form:
\be \label{eq:Snform}
\wt{S}^{(n)} \big(\{x\}\big) = \int \left(\prod_{i=1}^n \frac{d^4 k_i}{(2\pi)^4} \right) \ca{A}^{(n)} \big(\{k_s\}\big) \cM^{(n)} \big(\{x\}, \{k_s\}\big) \,,
\ee
where $\ca{A}^{(n)}$ is the $n$-loop squared matrix element and $\cM^{(n)}$ implements the measurement of the Fourier space observables $\{x\}$ on the $n$-particle final state.  Abelian matrix elements for $n$ particles factorize:
\be \label{eq:Afact}
\ca{A}^{(n)} \big(\{k_s\}\big) = \frac{1}{n!}\prod_{i=1}^n \ca{A}^{(1)} (k_i) \,,
\ee
where the one loop matrix element is
\be\label{eq:MEcf2}
\ca{A}^{(1)} ( k_i ) = 4g^2 C_F \, \frac{1}{(n\cdot k_i )( \bn\cdot k_i)} \, 2\pi \delta(k_i^2) \theta(k_i^0) \,.
\ee
For observables for which the measurement function factorizes, as in \eq{MeasPosAT}, the Abelian contributions to the soft function also factorize.  This implies Abelian exponentiation, where we use the common definition that the $n$-loop Abelian contribution to the soft function is determined solely by the one-loop contribution:
\begin{align} \label{eq:Sfact}
\wt{S}^{(n)}_{\text{Abel.}} \big(\{x\}\big) &= \frac{1}{n!} \left(\int \frac{d^4 k}{(2\pi)^4} \, \ca{A}^{(1)} (k) \, \cM^{(1)} \big(\{x\}, k\big)\right)^n \nn \\
&= \frac{1}{n!} \left[ \wt{S}^{(1)} \big(\{x\}\big) \right]^n \,,
\end{align}
which implies
\be
\wt{S}_{\text{Abel.}} \big(\{x\}\big) = \exp\Big[ \wt{S}^{(1)} \big(\{x\}\big) \Big] \,.
\ee

The renormalized one-loop soft function in Fourier space is (for all algorithms in the $\kt$ class)
\begin{align} \label{eq:S1loop}
\wt{S}^{(1)}(x,y) 
&= \frac{\as C_F}{\pi} \frac{1}{Q} \bigg[ -2 \ln^2 \left( \frac{\mu}{Q}\tan\frac{R}{2} \, ie^{\gamma_E} x \right) + 2 \ln \left( \tan^2 \frac{R}{2} \right) \ln\left( \frac{\mu}{2Q} \, ie^{\gamma_E} y \right) \nn \\
& \qquad\qquad\qquad - \frac12\ln^2 \left( \tan^2 \frac{R}{2} \right) - 2 \Li_2 \left( - \tan^2 \frac{R}{2} \right) - \frac{5\pi^2}{12} \bigg] \,.
\end{align}

\subsection{Violation of Abelian Exponentiation}
\label{ssec:Violabelexp}

Corrections in the soft function due to boundary clustering, $\Delta \wt{S}_\alg$, violate Abelian exponentiation, since the $n$-loop Abelian contribution to the soft function is no longer determined by the one-loop contribution. This effect starts at $\ca{O}(\as^2)$ and the relevant contributions at this order are shown in \fig{CAps}. Such configurations give logarithms that are sensitive to both $\rho$ and $\Lambda$ in momentum space, as we shall see by explicit calculation at $\ca{O}(\as^2 C_F^2)$ in \sec{AbelianNGLs}. In general each final state gluon can contribute a double logarithm to the cross section, $\as^n \ln^{2n} (Q \rho/\Lambda)$, with the double log associated with soft and collinear divergences. Unlike the non-Abelian case, the $\ca{O}(\as^n)$ Abelian soft matrix element, \eqs{Afact}{MEcf2}, has no $1/k_i\cdot k_j$ structures and so the only collinear divergences in the matrix element arise for soft gluons collinear to the jet directions. 

However, in order for gluons to boundary cluster and contribute to $\Delta S_\alg$ they cannot be collinear to the jet direction.  For C/A the closest a soft gluon can be to the jet axis and be merged across the jet boundary is $2^{-(n-1)}R$ for $n$ final state gluons.  For $\kt$, the condition is more subtle.  
\begin{figure}[t]{
	\begin{center}
	\includegraphics[width=.32\textwidth]{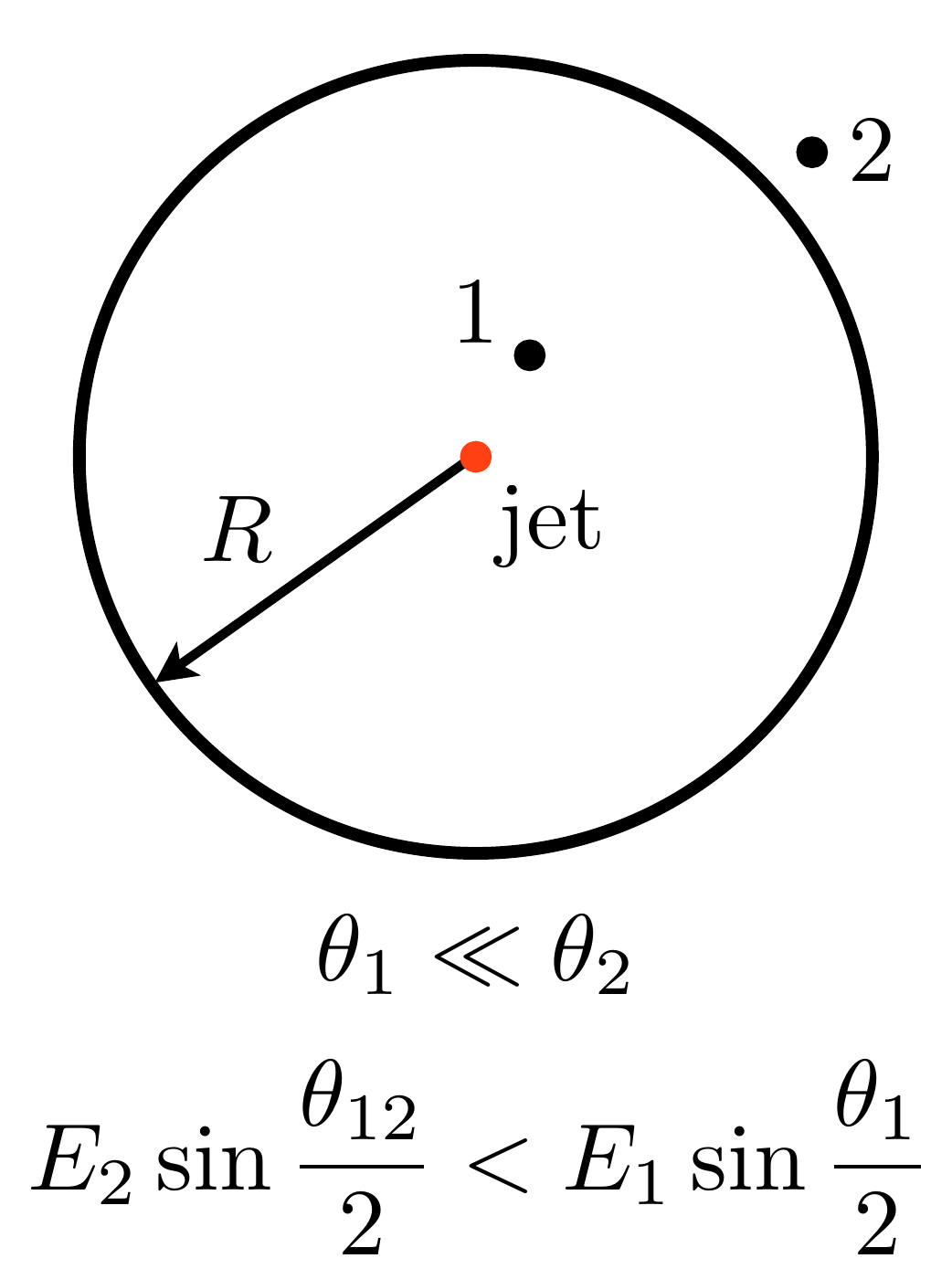} 
	  \end{center} 
	  \vspace{-1em}
{ \caption[1]{A configuration of two soft gluons that will boundary cluster.  Even though gluon 1 is close to the jet axis, gluon 2 is sufficiently soft so that it will cluster with gluon 1 before gluon 1 clusters with the jet.}
  \label{fig:kTconfig}} }
\end{figure}
Suppose we have two soft gluons 1 and 2, with gluon 1 at a small angle $\theta_1$ to the jet axis, and gluon 2 at a larger angle $\theta_2$ to the jet axis ($\theta_1 \ll \theta_2$).  If the two soft gluons cluster instead of gluon 1 clustering with the jet, then
\be
E_2 \sin \frac{\theta_{12}}{2} < E_1 \sin \frac{\theta_1}{2} \,.
\ee
Such a configuration is shown in \fig{kTconfig}.  Since gluon 1 is close to the jet axis, $\theta_1 \ll \theta_{12}$, meaning gluon 2 must be much softer than gluon 1:
\be
\frac{E_2}{E_1} < \frac{\sin \theta_1/2}{\sin\theta_{12}/2} \,.
\ee
This means that the combined pair will be nearly collinear with gluon 1.  If gluon 2 is outside of the jet, such that the clustering contributes to $\Delta S_\alg$, then since $\theta_{12} < R$ we must have
\be
R < \theta_{2} < R + \theta_1 \,.
\ee
This is a narrow range of angles, meaning the phase space for gluon 2 is power suppressed.  Although these configurations will contribute to the clustering logs at leading power, the power suppression of phase space for gluon 2 mitigates the collinear logarithmic enhancements from soft gluon clustering with $\kt$.  Since the momentum of a soft gluon close to the jet axis is nearly unchanged if it clusters with another soft gluon instead of the jet, this argument extends to configurations with more soft gluons.

For the Abelian terms for both the C/A and $\kt$ algorithms, we see that if a soft gluon contributes to $\Delta S_\alg(\rho,\Lambda)$ then it can contribute at most a single log from a soft divergence, since collinear divergences with the jet direction are excluded by boundary clustering. This provides a very straightforward log counting of the Abelian terms in $\Delta S_\alg$: for $n$ final state gluons, the leading term in $\Delta S_\alg^{(n)}$ is  at least of order $\as^n \ln^n(Q\rho/\Lambda)$, which is NLL.

\section{Abelian Clustering Non-Global Logarithms for the Dijet Mass Distribution}
\label{sec:AbelianNGLs}

In this section we calculate the $\ca{O}{(\as^2)}$ violations to Abelian exponentiation from soft clustering. We focus on the C/A and $\kt$ algorithms, but the same effect occurs for any jet algorithm that does not satisfy \eq{MeasPosAT}. It is useful to express the measurement for a generic algorithm in terms of the anti-$\kt$ measurement function $\cM_{\akt}$ and a correction term as in \eq{MeasN}. This gives a corresponding correction to the soft function $\Delta \wt{S}_\alg$, as in \eq{wtSdiff}, which at $\ca{O}(\as^2)$ can be written as
\be
\wt{S}^{(2)}_\alg = \frac{1}{2} \left[  \wt{S}_{\akt}^{(1)}\right] ^2 + \Delta \wt{S}^{(2)}_\alg\,.
\ee
 The effect of correlations in the soft gluon phase space are expressed entirely by the soft function correction term. In this section we work in momentum space, where analogous expressions hold, and we calculate $\Delta S^{(2)}_\alg(\rho, \Lambda)$ for C/A and $\kt$.

\subsection{Abelian Non-Global Logarithms at $\ca{O}(\as^2)$}
\label{ssec:CA}

The Abelian soft function correction term at $\ca{O}(\as^2)$ due to clustering with an algorithm is
\be \label{eq:S2contrib}
\Delta S_\alg^{(2)} (\rho, \Lambda) = \frac{1}{2} \int \frac{d^4 k_1}{(2\pi)^4} \, \frac{d^4 k_2}{(2\pi)^4} \, \ca{A}^{(1)} (k_1)\ca{A}^{(1)} (k_2) \, \Delta\ca{M}^{(2)}_\alg (\rho, \Lambda; k_1, k_2)\,,
\ee
where $\ca{A}^{(1)}(k_{1,2})$ is given in \eq{MEcf2}. Note that in pure dimensional regularization, $\as^2 \, C_F^2$ terms with a virtual gluon are scaleless and hence zero -- these terms make the soft function IR finite.

The configurations which contribute to the correction terms in the measurement function, $\Delta \ca{M}^{(2)}_\alg$, are shown in \fig{CAps}. The solid line represents the anit-$\kt$ jet boundary of radius $R$ around the jet direction. The clustering of soft gluons in and out of the boundary is represented by \fig{CAps}(a) and (b), which we refer to as the in-out contribution (IO), while \fig{CAps}(c) shows two gluons out of the jet being clustered in to the jet, which we refer to as the out-out contribution (OO). It is useful to define the constraints requiring the soft gluons to be in a region of phase space where the algorithm produces a different cross section than the anti-$\kt$ algorithm as $\Theta_{\alg}^\io$ and $\Theta_{\alg}^\oo$ for the in-out and out-out contributions respectively. We can write these schematically as
\begin{align} \label{eq:ThetaClusAlg}
\Theta_{\alg}^\io &= \theta(k_1 \text{ in $n$ jet}) \, \theta(k_2 \text{ out of jets}) \, \theta(k_1, k_2 \text{ cluster}) \nn \\
\Theta_{\alg}^\oo &=\theta(k_1 \text{ out of jets}) \, \theta(k_2 \text{ out of jets}) \, \theta(k_1, k_2 \text{ cluster}) 
\end{align}
For each contribution \fig{CAps}(a)-(c), we must remove the contribution of the anti-$\kt$ measurement to that region of phase space included in $\ca{M}_{\akt}$ in the momentum space version of \eq{MeasN}. We can write the measurement function correction term as
\begin{align} \label{eq:Mschematic}
\Delta\ca{M}^{(2)}_\alg (\rho,\Lambda;k_1,k_2) &= 4 \, \Theta_{\alg}^\io \left[\ca{M}^\io_{\rho} +\ca{M}^\io_{\Lambda}+\ca{M}^\io_{a} \right] + 2 \, \Theta_{\alg}^\oo  \left[ \ca{M}^\oo_{\rho} + \ca{M}^\oo_{a} \right]    \,,
\end{align}
where the in-out contributions from \fig{CAps}(a) and (b) are 
\begin{align}\label{eq:MIO}
\ca{M}^\io_{\rho} &= \theta(k_1 + k_2 \text{ in $n$ jet}) \, \delta\big(\rho - (k_1^+ + k_2^+)/Q \big) \, \delta( \Lambda) \,,  \nn \\
\ca{M}^\io_{\Lambda} &= \theta(k_1 + k_2 \text{ out of jets}) \, \delta(\rho) \, \delta(\Lambda - (k^0_1 + k^0_2))  \,,  \nn \\
\ca{M}^\io_{a} &= - \, \delta\big(\rho - k_1^+/Q\big) \, \delta(\Lambda - k^0_2) \,.
\end{align}
$\ca{M}^\io_{\rho}$ is the contribution where the algorithm clusters the gluons into the jet, $\ca{M}^\io_{\Lambda}$ is the contribution where the algorithm clusters the gluons out of the jet, and $\ca{M}^\io_{a}$ removes the contribution from the anti-$\kt$ algorithm in the correction term. 
The out-out contributions from \fig{CAps}(c) are  
\begin{align}\label{eq:MOO}
 \ca{M}^\oo_{\rho} &= \theta(k_1+k_2 \text{ in $n$ jet}) \, \delta\big(\rho - (k_1^+ + k_2^+)/Q \big) \, \delta( \Lambda) \,, \nn\\
 \ca{M}^\oo_{a} &= - \, \theta(k_1+k_2 \text{ in $n$ jet}) \,  \delta\big(\rho - k_1^+ /Q \big) \, \delta( \Lambda-k^0_2) \,.
 \end{align}
As with the in-out contributions, $\ca{M}^\oo_{\rho}$ is the contribution where the algorithm clusters the gluons into the jet and $\ca{M}^\oo_{a}$ removes the contribution from the anti-$\kt$ algorithm.  Note the combinatoric factors in the measurement function arise because interchanging $k_1$ and $k_2$ produces an identical contribution to the cross section and since there are two jets, $k_1$ can be in either jet for the in-out contribution.

\begin{figure}[t]{
	\begin{center}
	\includegraphics[width=.7\textwidth]{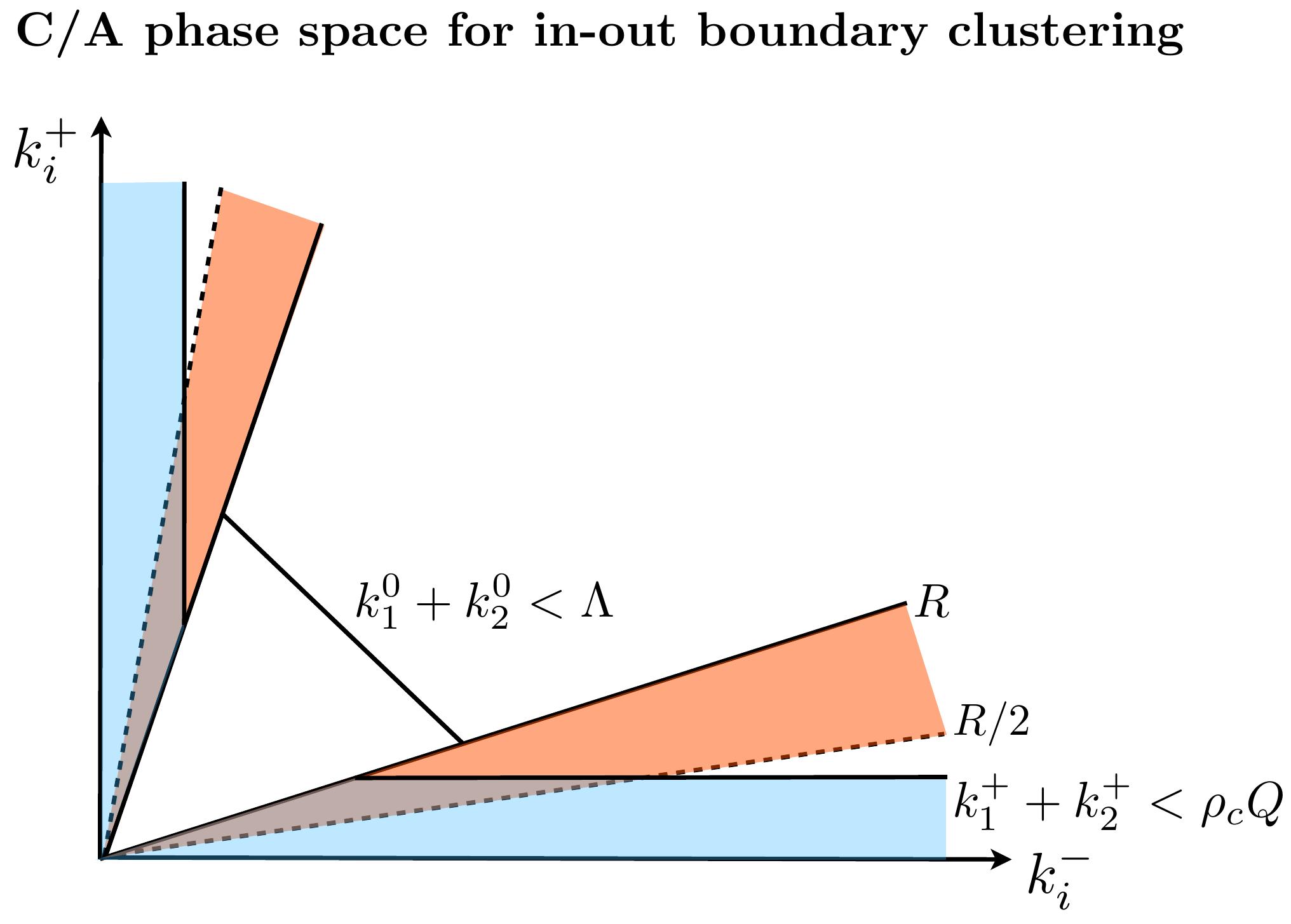} 
	  \end{center} 
	  \vspace{-1em}
{ \caption[1]{The C/A in-out region of phase space at $\ca{O}(\as^2)$ is shown schematically, where we overlay the phase space for gluon 1 and 2 in the $(k_i^+, k_i^-)$ plane. The region where the gluon in the jet can boundary cluster is shaded in red (dark gray). The region contributing to the jet mass cumulant $\rho_c$ is shaded in blue (light gray). The over lap between these two regions corresponds to $\ca{M}^\io_{\rho}$, the region outside the jets up to $k_1^0+k_2^0<\Lambda$ corresponds to $\ca{M}^\io_{\Lambda}$ and the combined regions are subtracted in $\ca{M}^\io_a$, removing the IR divergences from this contribution to the measurement function.   
}
  \label{fig:IOCAPS}} }
\end{figure}

We present the in-out regions of phase space for the C/A algorithm schematically in \fig{IOCAPS}. The region in red corresponds to gluon 1 in the jet, where boundary clustering can take place, while the region in blue contributes to the measurement of the jet mass cumulant $\rho_c$. The overlap between these two regions corresponds to $\ca{M}^\io_{\rho}$, the region outside up to $k_1^0+k_2^0<\Lambda$ contributes to $\ca{M}^\io_{\Lambda}$ and the combined regions are subtracted in $\ca{M}^\io_a$. For the Abelian terms, collinear divergences only arise along the axes, while soft divergences occur at the origin. It is clear from the figure that boundary clustering (red region) does not contain any collinear divergences and that the while each term in \eq{MIO} contributes soft divergences these are removed in the sum. 
We can see this explicitly by taking the soft limit of $\ca{M}_\alg^{(2)}$ in \eq{Mschematic}, which vanishes as both $k_1^0 \to 0$ and $k_2^0 \to 0$. \fig{IOCAPS} also illustrates that the jet mass measurement (blue region) constrains the phase space in the UV rather than the jet boundary $R$ and therefore the soft function anomalous dimension is independent of the algorithm.

\begin{figure}[t]{
	\begin{center}
	\includegraphics[width=.475\textwidth]{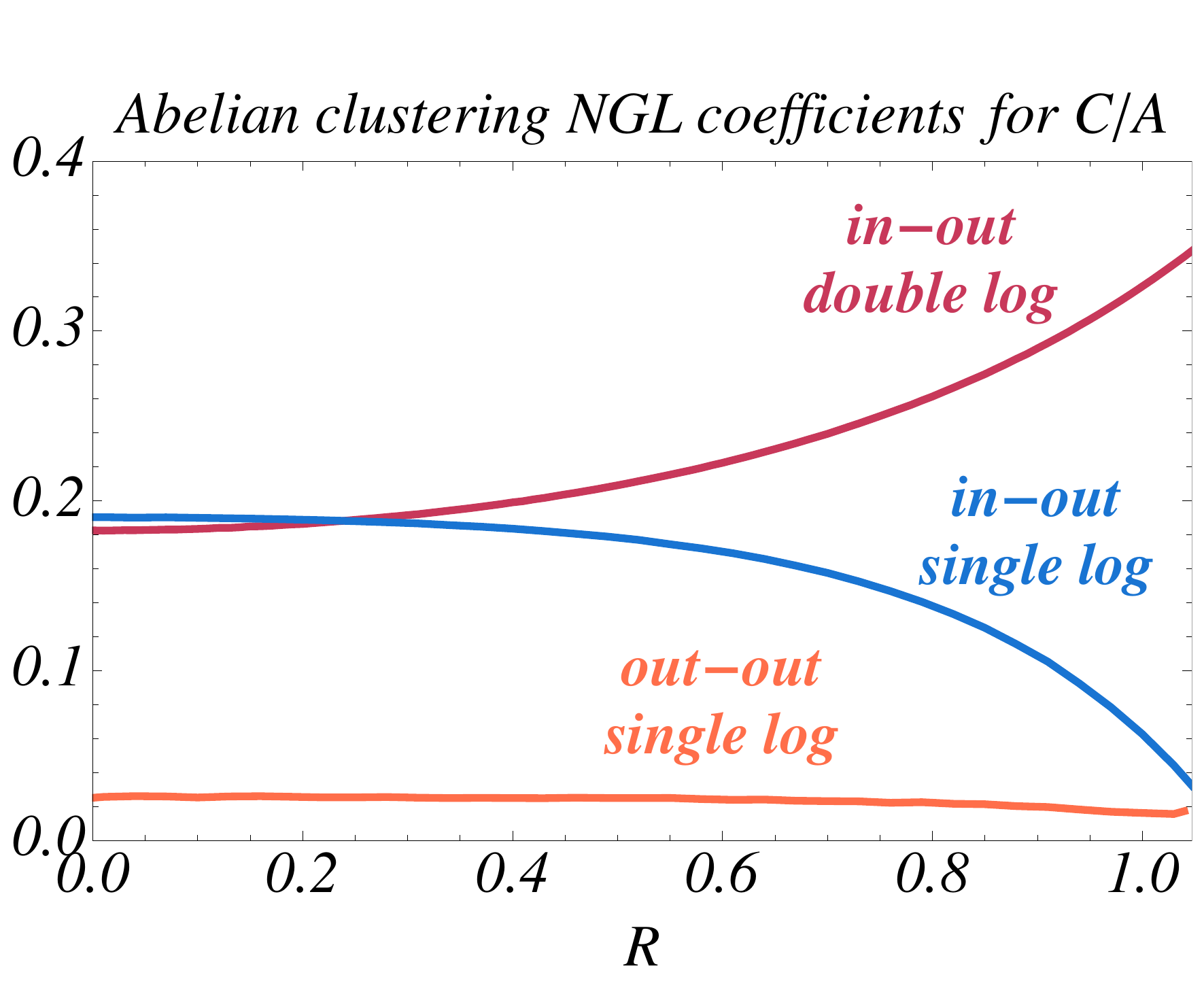} 
	\hspace{1em}
	\includegraphics[width=.475\textwidth]{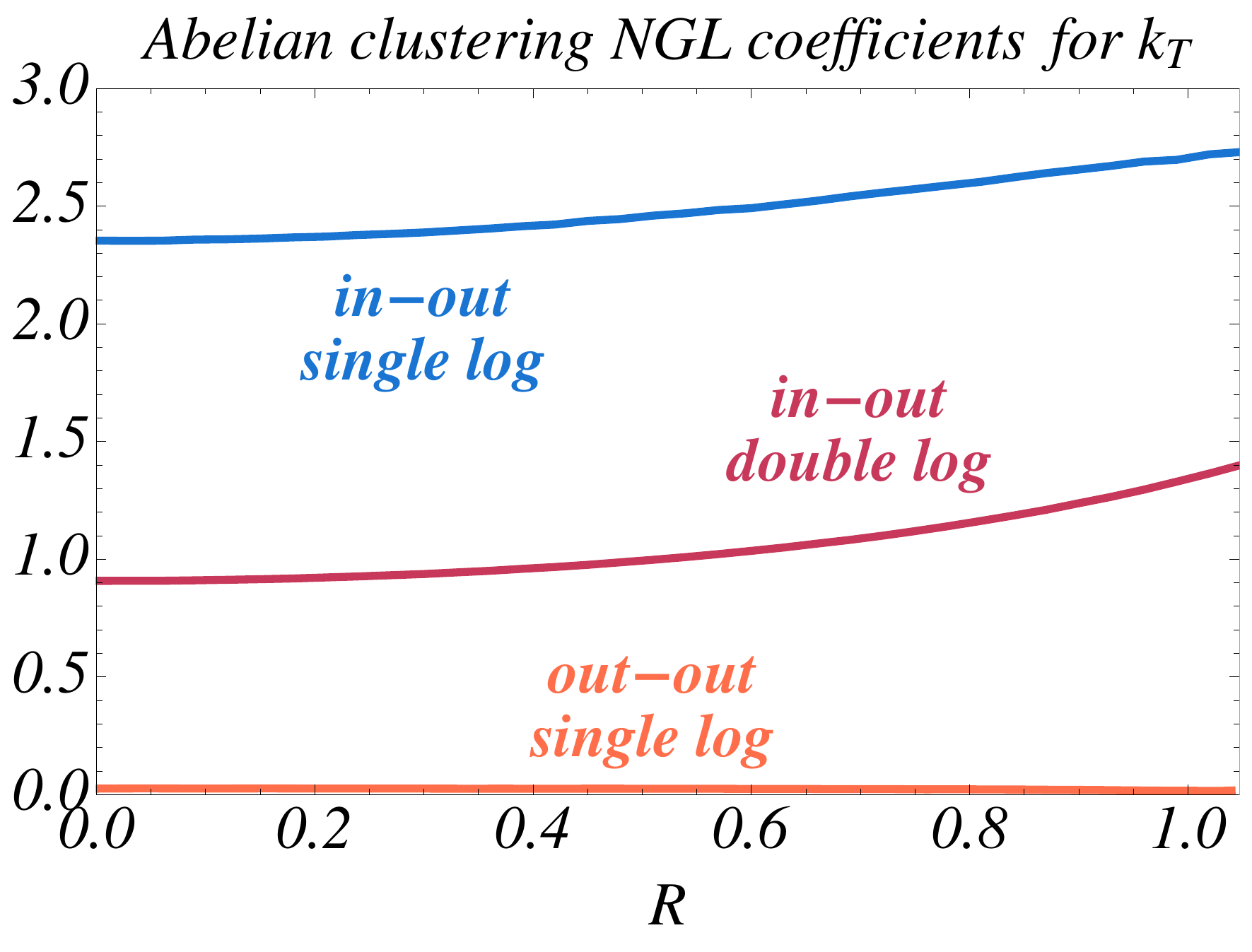}
	  \end{center} 
	  \vspace{-1em}
{ \caption[1]{The coefficients of the in-out double NGL, $C^{(0)}_{\alg} (R)$, the in-out single NGL, $C^{(1)}_{\io,\,\alg} (R)$, and the out-out single NGL, $C^{(1)}_{\oo,\,\alg} (R)$, as a function of $R$, for C/A (left) and $\kt$ (right).  These are the coefficients of $(\as C_F/\pi)^2 \ln^{2,1} Q\rho/ 2\Lambda \sin^2 \frac{R}{2}$ in the cumulant $\Delta\Sigma_S^{(2)}$.  For $\kt$, the in-out double and single log coefficients depend on the value of the argument of the NGLs, and we plot the coefficients for $Q\rho/ 2\Lambda \sin^2 \frac{R}{2} \ge 1$.  They are constant over this range.}
  \label{fig:NGLcoeffs}} }
\end{figure}

We present the results of the calculations of \eq{S2contrib} for C/A and $\kt$ here and leave the details to Appendix~\ref{sec:AppCalc}. As we have argued, the logarithms in $\Delta S_\alg(\rho,\Lambda)$ come from soft divergences, which we extract by integrating over $k^0_1$ and $k_2^0$. We express the final result in terms of coefficients of the double NGL ($C^{(0)}_{\alg} (R)$, which comes from in-out terms only), and the single NGL ($C^{(1)}_{\io,\,\alg} (R) + C^{(1)}_{\oo,\,\alg} (R)$), which we calculate numerically and plot in \fig{NGLcoeffs}.  The integrals that determine these coefficients are given in \eqs{finiteAs}{ADelta}.  For the $\kt$ algorithm, these coefficients depend on the values of $\rho$ and $\Lambda$.  This makes it difficult to write $\Delta S^{(2)}_{\kt}$ in terms of standard distribution functions, and so for the $\kt$ algorithm we only determine the cumulant.

For the C/A algorithm, the correction terms to the soft function due to correlations introduced by clustering are 
\begin{align}
&\Delta S_\CA^{(2)}(\rho,\Lambda) = \left( \frac{\as C_F}{\pi} \right)^2 \frac{1}{2Q\sin^2 \frac{R}{2}} \nn \\
& \quad \times \bigg\{  2 C^{(0)}_{\CA}(R) \bigg[ \delta\left(\frac{\rho}{2\sin^2 \frac{R}{2}}\right) \ca{L}_1 \left(\frac{\Lambda}{Q} \right)+ \delta\!\left(\frac{\Lambda}{Q} \right) \ca{L}_1\left(\frac{\rho}{2\sin^2 \frac{R}{2}}\right) - \ca{L}_0\left(\frac{\rho}{2\sin^2 \frac{R}{2}}\right) \ca{L}_0\!\left( \frac{\Lambda}{Q} \right) \bigg]  \nn \\
& \qquad \quad + \Big(C^{(1)}_{\io,\,\CA} (R) + C^{(1)}_{\oo,\,\CA} (R) \Big)  \bigg[ \delta\! \left(\frac{\Lambda}{Q} \right) \ca{L}_0 \left(\frac{\rho}{2\sin^2 \frac{R}{2}}\right) - \delta\left(\frac{\rho}{2\sin^2 \frac{R}{2}}\right)\ca{L}_0 \left(\frac{\Lambda}{Q} \right) \bigg] \nn \\
& \qquad \quad + (\text{constant})\, \delta\left(\frac{\rho}{2\sin^2 \frac{R}{2}}\right)\, \delta\!\left( \frac{\Lambda}{Q} \right) \bigg\} \,.
\end{align}
$\ca{L}_0$ and $\ca{L}_1$ are standard plus distributions \cite{Ligeti:2008ac},
\be
\ca{L}_0 (x) = \bigg[\frac{\theta(x)}{x}\bigg]_+ \,, \qquad \ca{L}_1 (x) = \bigg[ \frac{\theta(x)\ln x}{x}\bigg]_+ \,.
\ee
The term proportional to $\delta(\rho)\delta(\Lambda)$ is a number dependent only on $R$ that we do not calculate.  For the cumulant we find the same form for both algorithms,
\begin{align} \label{eq:Scumulant}
\Delta \Sigma_S^{(2)} &= \left( \frac{\as C_F}{\pi} \right)^2 \bigg\{ C^{(0)} (R) \ln^2 \frac{Q\rho_c}{2\Lambda_c \sin^2 \frac{R}{2}} + \Big( C^{(1)}_{\io,\,\alg} (R) + C^{(1)}_{\oo,\,\alg} (R) \Big) \ln \frac{Q\rho_c}{2\Lambda_c \sin^2 \frac{R}{2}} \nn \\
& \qquad \qquad \qquad \qquad + \text{ constant}\, \bigg\} \,,
\end{align}
For the $\kt$ algorithm the in-out single and double log coefficients depend on the value of the argument of the log, $Q\rho_c/2\Lambda_c \sin^2 R/2$.  As discussed in Appendix~\ref{sec:AppCalc}, when this ratio is larger than 1 the coefficients are constant, and decrease as this ratio becomes smaller than 1.  In \fig{NGLcoeffs}, we plot the $\kt$ coefficients for the regime $Q\rho_c/2\Lambda_c \sin^2 R/2 \ge 1$.  In the next section we compare our results to the distribution calculated using the Monte Carlo program EVENT2.

The coefficient of the leading double logarithm of $\as^2 C_F^2 \ln^2 \rho_c$ in the correction term for the cumulant was calculated for C/A in the limit of small jet size, $R$, in \cite{KhelifaKerfa:2011zu}. We find our result for the double log coefficient for C/A in the limit of small $R$,
\be
\lim_{R\to0} C^{(0)}_{\CA}(R) = 0.183 \,,
\ee
agrees within the precision of the value in \cite{KhelifaKerfa:2011zu}, which is 0.185 in our units.

\subsection{Comparison to EVENT2}
\label{ssec:EVENT2}

We can compare the results of our calculations to the output of the Monte Carlo program EVENT2 \cite{Catani:1996jh,Catani:1996vz}.  EVENT2 implements Catani-Seymour subtractions and contains the matrix elements necessary to compute observables in $e^+e^-$ events to $\ca{O}(\as^2)$ that vanish in the 2-jet limit.  We use EVENT2 to calculate the dijet mass distribution, taking the cumulant in the out-of-jet energy cut $\Lambda_c$:
\be
\frac{d\sigma}{d\rho} (\Lambda_c) = \int_0^{\Lambda_c} d\Lambda \frac{d^2\sigma}{d\rho\, d\Lambda} \,.
\ee
EVENT2 calculates a binned distribution in $\rho$, with the cross section in a bin $[\rho_{\min}, \rho_{\max}]$ given by
\be
\sigma_{\text{EV2}} (\rho_{\min},\rho_{\max};\Lambda_c) = \int_{\rho_{\min}}^{\rho_{\max}} d\rho \, \frac{d\sigma}{d\rho} (\Lambda_c) = \Sigma (\rho_{\max}, \Lambda_c) - \Sigma (\rho_{\min}, \Lambda_c) \,,
\ee
where $\Sigma$ is the double cumulant of the cross section, \eq{cumulantdef}.
\begin{figure}[t!]{
 \includegraphics[width=.95\textwidth]{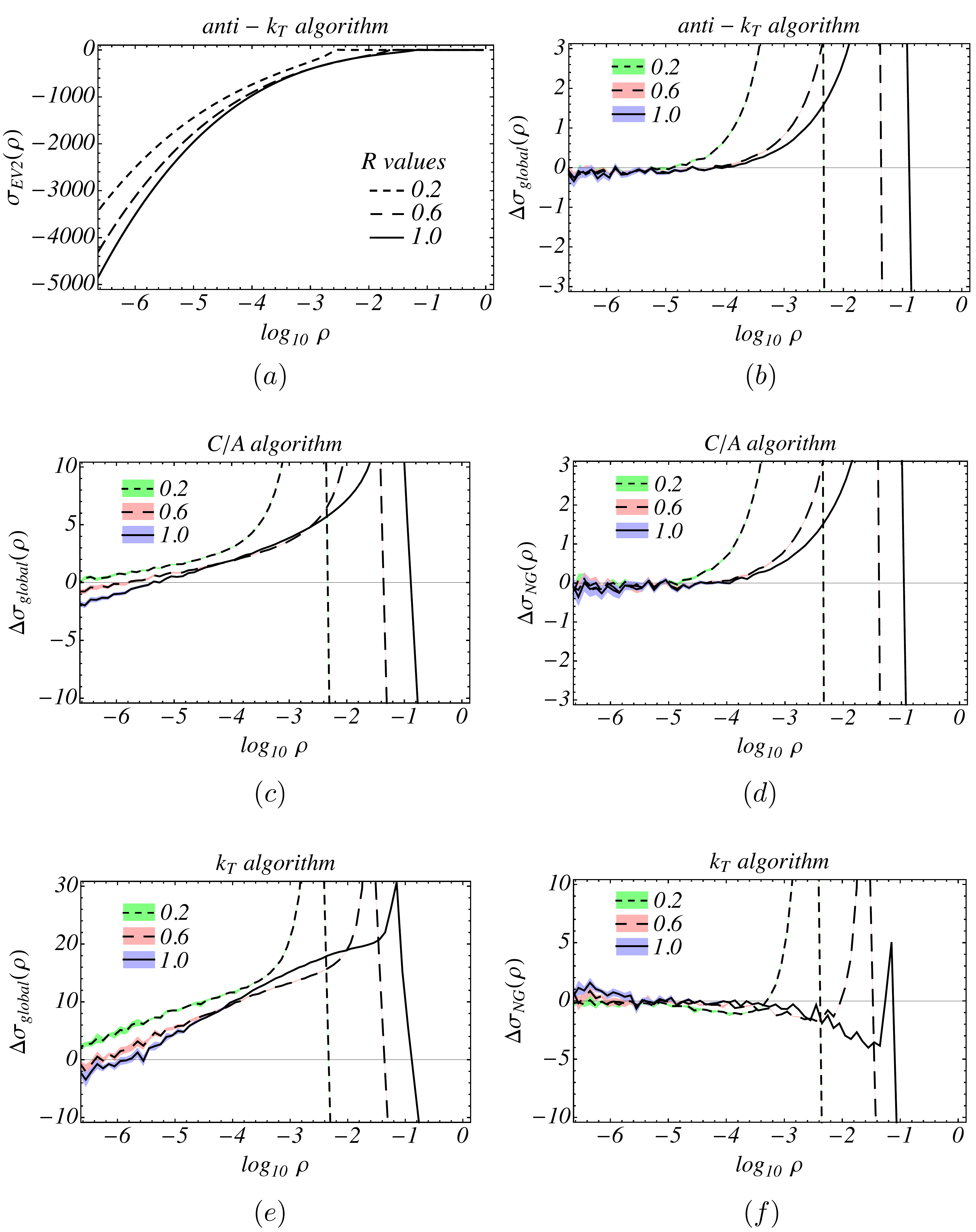}
{ \caption[1]{Numerical analysis of the EVENT2 results with $\Lambda_c/Q = 10^{-5}$.  In (a), the Abelian terms in the distribution $d\sigma/d\rho$ for the anti-$\kt$ algorithm are shown.  The distribution after the global logs have been subtracted is shown in (b).  The result tends to 0 for small $\rho$, verifying the global logs.  In (c) and (e), the distributions for the C/A and $\kt$ algorithms after the global logs have been subtracted are shown.  The distributions are not flat for small $\rho$, meaning double logarithms remain.  These are the clustering NGLs, which are subtracted in (d) and (f).  The distribution goes to 0 for small $\rho$, confirming the calculations of the clustering NGLs.}
\label{fig:Event2all}} }
\end{figure}

We compare the $\ca{O}(\as^2 C_F^2)$ terms to EVENT2, first subtracting the global logs from the distribution from EVENT2.  These global logs, equal to the anti-$\kt$ cross section, are given by the universal hard and jet functions and the anti-$\kt$ soft function through $\ca{O}(\as^2)$; for reference, the global logs are given in Section V of \cite{Hornig:2011tg}.  We compare the analytic results for the anti-$\kt$, C/A, and $\kt$ algorithms to EVENT2.  The anti-$\kt$ comparison serves to verify that the global logs are correct, and the C/A and $\kt$ comparisons allow us to directly test the calculations of the Abelian NGLs.  For each algorithm, we use EVENT2 to calculate $\sigma_{\text{EV2}}$ for $\Lambda_c/Q = 10^{-5}$ and $R =$ 0.2, 0.6, and 1.0.  We generated approximately $2\cdot 10^{12}$ events for the anti-$\kt$ and C/A algorithms and approximately $8\cdot 10^{12}$ events for the $\kt$ algorithm.  The $\kt$ results are less numerically stable, but still allow us to test the calculation of the NGLs.

In \fig{Event2all}, we show the distributions from EVENT2 that confirm the calculations of the clustering NGLs.  For the anti-$\kt$ algorithm, we show the distribution $d\sigma/d\rho$ before and after the global logs have been subtracted.  After the global logs are removed, the distribution vanishes for small $\rho$, indicating that only power suppressed terms in $\rho$ remain and verifying the global logs.  For the C/A and $\kt$ algorithms, we subtract the global logs and show the distribution $d\sigma/d\rho$ before and after the clustering NGLs have been subtracted.  Before the clustering NGLs are subtracted, the distributions for small $\rho$ have a non-zero slope, indicating double logs remain.  After they are subtracted, the distribution vanishes for small $\rho$, confirming our calculations.

\section{Non-Global and Clustering Logarithms}
\label{sec:NGL}
In this section we discuss general properties of clustering logs that we saw arise explicitly in the $\ca{O}(\as^2)$ calculation in the previous section. Clustering logs share many of the same properties as traditional NGLs, leading us to think of them as a class of NGLs (clustering NGLs).  We will discuss the points of similarity outlined in the introduction in further detail.  

In order for clustering logs to arise, two properties must be satisfied. First, there must be correlations in the phase space constraints on multiple gluons in the final state, requiring that $\Delta \cM$ in \eq{MeasN} is non-zero. Second, the measurement must be divided into regions with different scales contributing to the cross section, namely $\Delta \cM$ must depend on multiple scales.  These conditions are satisfied by boundary clustering, which requires at least one particle to move between two measurement regions and gives rise to clustering logs of the ratio of scales associated with these regions.  

The conditions required for traditional NGLs to arise are completely analogous.  There must be a correlation between the soft gluons from the non-Abelian matrix elements and there must exist a boundary in phase space where the measurements differ on opposite side of the boundary.  The latter is a feature of the observable, and many jet observables of interest contain NGLs. A typical configuration that gives rise to NGLs is shown in \fig{NGLdiagram}(b). NGLs have been studied for a variety of observables at $\ca{O}(\as^2)$. For the distribution we study here, $d^2\sigma/ d\rho \, d\Lambda$, the non-Abelian NGLs are\footnote{The power of $\tan\frac{R}{2}$ in \eq{NGLform} is different from that in \cite{Hornig:2011tg}. Since we are not working in the small $R$ limit, both represent valid choices. With the scaling of $R$ included in the argument of the NGLs in this paper, the coefficients of the $\as^2 \ln^2$ and $\as^2 \ln$ terms are finite as $R\to0$.} 
\be\label{eq:NGLform}
\left[f_{\akt}+\Delta f_{\alg}\right] \left(\frac{\as}{2\pi}\right)^2 C_F C_A \, \ln^2 \frac{Q\rho}{2 \Lambda \tan^2 \frac{R}{2}} \,.
\ee
The correction due to clustering in the measurement function, $\Delta \ca{M}_{\alg}$, gives rise to a non-Abelian clustering NGL with a coefficient $\Delta f_{\alg}$ that depends on the jet algorithm.  For the anti-$\kt$ and cone algorithms, for which clustering logarithms do not contribute, $\Delta f_{\akt}$ is power suppressed. The coefficient $f_{\akt}$ can be determined analytically \cite{Hornig:2011tg,Kelley:2011tj}, and is
\be
f_{\akt} = -\frac{2\pi^2}{3} + 4 \Li_2 \left( \tan^4 \frac{R}{2} \right) \,.
\ee
The contribution of non-Abelian clustering NGLs, $\Delta f_{\alg}$, has been calculated for C/A and $\kt$ numerically in \cite{Hornig:2011tg}. These reduce the magnitude of the coefficient in \eq{NGLform} for algorithms where soft gluon clustering plays a role.

We saw from the the argument in \ssec{Violabelexp} and the explicit calculation in \sec{AbelianNGLs} that clustering NGLs arise from soft singularities. This is true for both Abelian and non-Abelian clustering NGLs. When boundary clustering merges a particle in the jet with one out of the jet, the particle in the jet cannot be collinear to the jet axis and hence cannot be sensitive to the collinear singularity.  This is required by a collinear safe jet algorithm: if a soft particle is collinear with the jet axis, it must cluster with the jet.  At leading power, there will be some region around the jet axis where boundary clustering cannot take place.  This implies that clustering NGLs arise from soft divergences rather than collinear. 

Traditional NGLs also arise from soft divergences. For non-Abelian real emission terms, in addition to the singularities corresponding to the partons becoming soft or collinear to the jet direction, there is an additional collinear singularity when the two partons align with each other. This singularity generates logarithms of a single scale if the the partons  propagate into the same region, or two scales if they propagate into different regions. The relationship between these terms contributing to the NGL was explored in \cite{Hornig:2011iu,Hornig:2011tg}.  For configurations with partons in different measurement regions (such as \fig{NGLdiagram}(b)), the collinear singularity is never realized due to the boundary.  A collinear enhancement remains in the matrix element, and the region of phase space with the two soft partons near the boundary dominates the value of the leading NGL coefficient.

Finally, we can show that clustering NGLs are not associated with any UV or IR divergences in the total soft function.  The lack of IR divergences follows directly from the fact that anti-$\kt$, C/A, and $\kt$ are IR safe jet algorithms.  The difference in their contribution to the soft function, and the cross section, cannot contain any IR divergences.  Since boundary clustering does not occur for the anti-$\kt$ algorithm, boundary clustering cannot give IR divergences for any IR safe algorithm.  The lack of UV divergences follows from the consistency condition on the factorization theorem in \eq{sigmafact}.  Consistency requires
\be\label{eq:consistency}
\gamma_H (Q)+ \gamma_{J_n}(\sqrt{\rho} \, Q)+ \gamma_{J_{\bar{n}} } (\sqrt{\rho} \,Q)+ \gamma_S (\rho \,Q)= 0 \,.
\ee
For the dijet mass distribution, when $\rho \ll R^2$ the jet function is inclusive and is independent of the choice of algorithm.  This is true for other observables where the phase space constraints in the far ultraviolet are more restrictive than the jet algorithm \cite{Bauer:2011hj}.  In this case, the hard function and jet function anomalous dimensions are independent of the jet algorithm, and \eq{consistency} requires that the soft function anomalous dimension is independent of the jet algorithm as well.  As far as RGE is concerned, there is no difference between the anti-$\kt$, C/A, and $\kt$ algorithms.  This means that clustering effects cannot contribute to the soft anomalous dimension, and hence cannot contribute UV divergences to the soft function or the cross section.

Similarly, traditional NGLs are independent of the UV divergences in the soft anomalous dimension. Since the hard and jet function anomalous dimensions are independent of the scale $\Lambda$, \eq{consistency} implies that the soft anomalous dimension is independent of $\Lambda$ as well. The non-Abelian NGL involving the scale $\Lambda$ in \eq{NGLform} must therefore be independent of the renormalization scale $\mu$ and not contribute the soft function anomalous dimension. The details of this structure are described in \cite{Hornig:2011tg}.

\section{Non-Global Logarithms at All Orders}
\label{sec:resum}

Having confirmed the existence of clustering NGLs and established their general properties and connection to traditional NGLs, we discuss the general structure of all NGLs beyond $\ca{O}(\as^2)$, focusing on the implications for resummation. Resummation enforces a relationship among the coefficients in the perturbative expansion of the soft function at a given order in an $\as^n \ln^m$ counting scheme. Consider for example the perturbative structure of the cumulant soft function. The perturbative series can be arranged according to log counting in the exponent as follows
\begin{align}
\Sigma_s =\exp \left[  \tL \, g_0(\as \tL) +g_1(\as \tL) +\as g_2(\as \tL) +\cdots \right]
\end{align} 
where $\tL$ represent logs of $\rho$ or $\Lambda$. The LL series is given by $\tL\, g_0(\as \tL) = d_{12} \,\as \tL^2+d_{23} \, \as^2 \tL^3+d_{34} \,\as^3\tL^4 +\cdots$, the NLL series is given by $g_1(\as \tL) = d_{11}\as \,\tL +d_{22} \, \as^2 \,\tL^2+\cdots$ and the NNLL series by $\as g_2(\as \tL) = d_{10} \,\as+d_{21} \, \as^2\, \tL+\cdots$. Resummation at order $\text{N}^k\text{LL}$ in the exponent allows all $g_i$ for $i \le k$ terms to be included.

Since clustering NGLs arise due to correlations in soft gluon phase space introduced by the jet algorithm, the specific structure of clustering logs at higher orders depends on the algorithm's metric and the number of gluons in the final state.  As we have seen, the $\kt$ class of jet algorithms have the same single soft gluon phase space constraints and therefore agree at $\ca{O}(\as)$.  However, they disagree when two or more gluons are in the final state, starting with the coefficient $d_{22}$ in the NLL series for both the Abelian and non-Abelian terms. We expect this disagreement to continue at higher orders.  For example,  the clustering constraints at $\ca{O}(\as^3)$ include the region
\be
\label{eq:BC3gluons}
\Delta \cM^{(3)}_\alg \supset \,  \theta\left( \text{gluons 1,2 and 3 boundary cluster}\right) \,, 
\ee
which arises first at this order and is unrelated to the clustering effects at $\ca{O}(\as^2)$ for two soft gluons.  As demonstrated in \ssec{Violabelexp}, the constraint of clustering means this contribution will be $\ca{O}(\as^3\tL^3)$, thus it adds to the $d_{33}$ term in the Abelian NLL series. Boundary clustering of $n$ gluons prevents collinear divergences with the jet direction; thus Abelian clustering NGLs arise at order $\as^n {\tL}^n$.  When only a subset of the soft gluons are clustered, the Abelian clustering NGLs could enter at order $\as^n {\tL}^m$, where $m>n$.  As a good example, again consider the $\Delta \cM^{(3)}_\alg$ correction, which also includes the region
\begin{align}\label{eq:2clus1not}
\Delta \cM^{(3)}_\alg \supset & \,  \theta\left( \text{gluon 1 is in the jet and does not boundary cluster}\right) \nn\\
& \times \theta\left( \text{gluon 2 and 3 boundary cluster}\right) \,.
\end{align}
This will contribute at order $\as^3 \tL^4$ to the Abelian terms because gluon $1$ is sensitive to both soft and collinear divergences and so can contribute ${\as \tL^2}$ to the cumulant, while gluons $2$ and $3$ are sensitive to soft divergences only and therefore contribute $\as^2 \tL^2$. This will provide a new contribution from clustering to the LL series coefficient $d_{34}$ unless the phase space constraints in \eq{2clus1not} can be expressed in terms of lower order constraints. That is, if gluon $1$ gives $d_{12} \as \tL^2$ and gluons $2$ and $3$ give the same contribution from boundary clustering at 2 loops, $d_{22} \as^2 \tL^2$, the clustering effects from \eq{2clus1not} will still be NLL. This is non-trivial to show and we leave the proof to future work \cite{KWZinprepMeas}, where we use a novel framework to express the all-orders form of the measurement function and use this to show that Abelian clustering NGLs contribute only at order $\as^n \tL^n$ for all $n \ge 2$.

This presents a troubling picture for resummation of clustering NGLs. Unless the phase space constraints giving the leading clustering NGLs for $n \geq 3$ particles are related to lower order phase space constraints for some algorithm, the coefficients in the NLL series in the exponent will be unrelated. There is no \textit{a priori} reason to believe that such a property exists, and considering the new Abelian clustering contributions at $\as^2$ and from \eq{BC3gluons} at $\as^3$, such a relationship between phase space constraints seem very unlikely.  This is further supported by the general structure of the measurement function that we prove in \cite{KWZinprepMeas}. We therefore find that the NLL resummation of clustering NGLs is likely impossible.  This is not necessarily true for traditional non-Abelian NGLs, as we discuss shortly.

We note that, even though resummation is likely broken at NLL,  we might be interested in log counting in the distribution instead instead of the exponent.  In this case, NLL means resumming all terms through order $\as^n \ln^{2n-2}$;  therefore, it is sufficient to perform leading log resummation in the exponent and then add the fixed order contribution from the clustering NGLs.

Non-Abelian NGLs arise from correlations in the matrix element for multiple soft gluon production.  These NGLs are determined by the perturbative structure of matrix elements of the soft function, which is the eikonal limit of QCD.  Unlike NGLs from soft gluon clustering, there is reason to believe that NGLs at all orders may be related.  The fundamental problem with resumming clustering NGLs is that the action of the jet algorithm at higher orders involves phase space constraints that are not present at lower orders.  Eikonal matrix elements at higher orders in QCD are related to lower order matrix elements, especially when organized by log counting in the exponent of the soft function.  

Consider an observable, such as thrust ($\tau$), that does not have NGLs.  Like the anti-$\kt$ algorithm, the measurement function for thrust is a product of single particle constraints.  Resummation can be performed in the exponent of the distribution in this case.  For thrust, this has been performed to $\text{N}^3\text{LL}$ \cite{Becher:2008cf}.  For observables of this type, resummation of a fixed order calculation includes contributions from higher order matrix elements that contribute at the same order in log counting.  The fact that resummation can be performed implies that, unlike the case of clustering, there is a relationship between the logarithmic contributions of matrix elements at all orders.  Successful resummation of non-Abelian NGLs would rely on this structure.  Currently the leading NGLs have been resummed in the large $N_c$ limit, where such all orders relationships between matrix elements are used \cite{Banfi:2002hw}.

It is expected that NGLs will contribute at $\text{NLL}$ in the exponent at all orders.  For the anti-$\kt$ algorithm, the measurement function for the dijet mass distribution factorizes into single gluon constraints, meaning that only correlations in the matrix element give rise to the NGLs in this observable.  The challenge in resumming such NGLs can be phrased in terms of understanding how the matrix elements that give rise to the higher order NGL terms are related to the lower order matrix elements.  This is well understood for observables such as thrust, as discussed above, but no such framework exists for NGLs.  If this framework can be developed, it may prove applicable for resummation of a wide range of observables.

\section{Conclusions}
\label{sec:conclusions}

Clustering due to jet algorithms can introduce correlations into soft gluon phase space constraints and ruin the simple picture of resummation, spoiling even Abelian exponentiation. This effect has been known for some time \cite{Catani:1991gn} and studied in specific cases \cite{Banfi:2005gj,Delenda:2006nf, KhelifaKerfa:2011zu}.  On the other hand, NGLs are traditionally thought of as arising from correlations in the soft gluon matrix element for measurements in separate regions of phase space, occurring in the non-Abelian terms first at $\as^2 C_F C_A \ln^2 (Q\rho/\Lambda)$.  Considering clustering logarithms in this light, we found it natural to interpret them as a class of NGLs.  Like traditional non-Abelian NGLs, they arise in the limit of gluons becoming soft rather than collinear.  They are not associated with UV divergences, and therefore do not effect the anomalous dimension of the matrix elements in the factorized cross section and consequently are not resummed by standard RGE.

Correlations in the measurement function arise in a wide range of jet observables.  We considered a straightforward observable to highlight this effect, the $e^+ e^- \to 2$ jet cross section with measurements of the total jet mass $\rho$ and the energy outside the jet $\Lambda$ for the anti-$\kt$, C/A, and $\kt$ jet algorithms. We worked in the limit $\{Q \rho, \Lambda\} \ll Q \sqrt{\rho} \ll Q$, where SCET provides the relevant description, and took the jet size $R \gg \sqrt{\rho}$. In this regime the soft function contains all of the nontrivial jet algorithm dependence.  In \sec{dijet}, we showed that the anti-$\kt$ jet algorithm obeys Abelian exponentiation and gave its all-orders form. This is because the anti-$\kt$ algorithm does not introduce correlations in soft gluon phase space constraints, which has been shown to be true to leading order in SCET power counting \cite{Walsh:2011fz}. We also examined the violation to Abelian exponentiation due to the C/A and $\kt$ algorithms, using constraints from the boundary clustering phase space regions to show that for $n$ final state gluons, the leading clustering logs are at least of order $\ca{O}(\as^n \ln^n Q\rho/\Lambda)$, which is NLL. 

In \sec{AbelianNGLs}, we calculated the Abelian soft function at $\ca{O}(\as^2)$ up to constant terms for the C/A and $\kt$ algorithms, including the full $R$ dependence of the coefficients of the logarithms. We verified our results of the leading and next-to leading Abelian NGLs are in agreement with the distribution $d\sigma/d\rho$ from the Monte Carlo program EVENT2. Taking the small $R$ limit for the leading clustering logarithm for C/A, we reproduce the result of \cite{KhelifaKerfa:2011zu}.  Analogous clustering NGLs exist for other factorizable jet algorithms, with calculable coefficients.

We outlined the properties of clustering NGLs in \sec{NGL} and compared these to traditional NGLs, showing that their key features are the same when examined in the context of SCET.  In \sec{resum}, we discussed the all-orders structure of NGLs.  We argued that at each order $\as^n$, there is a new correction term $\Delta \wt{S}^{(n)}$ that comes from clustering effects of $n$ soft particles contributing at least at NLL.  We show that this is the case in later work \cite{KWZinprepMeas} by deriving an all orders form of the measurement function, which allows us to prove the all-orders form of the Abelian soft function. This has serious implications for resummation.  Since each correction term $\Delta \wt{S}^{(n)}$ for $n\ge2$ contributes at least at NLL and the coefficients for different $n$ appear to be unrelated, it seems impossible to resum the Abelian contributions to the cross section at NLL using traditional log counting in the exponent of the distribution.  We also discussed the prospect for resummation of traditional NGLs.  These NGLs are based on the perturbative structure of eikonal matrix elements in QCD.  Since resummation exploits relationships between the logarithmic structure that arises from these matrix elements, it is likely that a similar relationship exists for non-Abelian NGLs.

\section{Acknowledgments}

We would like to thank Christopher Lee and Mrinal Dasgupta for useful conversations, and we appreciate comments on a draft of this work from Christian Bauer, Christopher Lee, and Matthew Schwartz.  This work was supported in part by the Office of High Energy Physics of the U.S.\ Department of Energy under the Contract DE-AC02-05CH11231 and Grant DE-SC003916. JW was supported in part by a LHC Theory Initiative Postdoctoral Fellowship, under the National Science Foundation grant PHY-0705682.

\appendix

\section{Calculation of $\ord{\as^2}$ Abelian Non-Global Logarithms for the C/A and $\kt$ Algorithms}
\label{sec:AppCalc}

In this appendix we give the details of the $\ca{O}(\as^2)$ calculations of the Abelian non-global logarithms in the C/A and $\kt$ dijet distributions in the sum of the jet masses, $Q \rho$, and energy outside the jets, $\Lambda$. To determine the violation to Abelian exponentiation we focus on $\Delta \ca{M}^{(2)}_\alg$ in \eq{MeasN}, which corresponds to the region of phase space where the algorithm clustering differs from the anti-$\kt$ algorithm and gives rise to a change in the cross section from anti-$\kt$.

The Abelian $\as^2$ soft function correction term due to clustering is given in \eq{S2contrib}. Recall Abelian contributions (in this case to $\as^2 C_F^2$) with virtual gluons are scaleless in pure dimensional regularization and so we do not consider them here.  The phase space constraints and matrix element only depend on the light cone coordinates $k_1^{\pm}, k_2^{\pm}$, and the angle $\phi$ between $k_1$ and $k_2$ in the plane transverse to the jet axes.  Therefore we can perform integrals over the other momentum components using the on-shell conditions.  In pure dimensional regularization with $d = 4-2\e$ dimensions in the $\overline{\text{MS}}$ scheme, the soft function correction term is
\begin{align}
\Delta S_\alg^{(2)} (\rho, \Lambda) &= 2 \left( \frac{\as C_F}{2\pi} \right)^2 \frac{\big(\mu^2 \, e^{\gamma_E}\big)^{2\e}}{\Gamma(1-\e)^2} \,\left(\frac{\pi^{1/2} \Gamma(\frac12 - \e)}{\Gamma(1-\e)}\right)^{-1} \int_0^{\pi} d\phi \, \sin^{-2\e} \phi \nn \\
& \qquad \times \int_0^{\infty} dk_1^+ \, dk_1^- \, dk_2^+ \, dk_2^- \, (k_1^+ k_1^- k_2^+ k_2^-)^{-1-\e} \, \Delta\ca{M}^{(2)}_\alg (\rho, \Lambda; k_1, k_2) \,.
\end{align}
The calculation of the correction terms are simplest using the coordinates of the energy and angle of the gluons,
\be
k^+_i \equiv k_i^0(1-\cos \theta_{i}) \,, \quad k^-_i \equiv k^0_i (1+\cos \theta_{i})\,, \quad \Rightarrow \quad dk^+_i dk^-_i = 2 \,k^0_i \, dk^0_i \, d\cos\theta_i \,,
\ee
where the angle $\theta_{i}$ is measured with respect to the $\hat{n}$ direction. Since the clustering of soft gluons across the jet boundary is free of collinear divergences with the jet direction, by IR safety of the jet algorithm, divergences will arise in the measurement function $\Delta\ca{M}^{(2)}_\alg $ in the $k^0_1$ and $k^0_2$ integrals. It is simpler to extract these divergences if we change the $u$ coordinates to $u_t$ and $z$,
\be
k^0_t \equiv k^0_1 + k^0_2 \,, \quad z \equiv \frac{k^0_2}{k^0_1 + k^0_2} \,, \quad \Rightarrow \quad dk^0_1 dk^0_2 = k^0_t \, dk^0_t dz \,.
\ee

\subsection{The Measurement Functions for the Correction Terms}

The terms that contribute to $\Delta \ca{M}^{(2)}_\alg$ are shown in \fig{CAps} and given schematically in Eqs.~\ref{eq:Mschematic},~\ref{eq:MIO}, and~\ref{eq:MOO}.  We now express the measurement functions in terms of the variables that are convenient for the calculation of the correction terms, $\{z, k_t^0,\cos \theta_1,\cos \theta_2,\cos \phi \}$.

For C/A, we can write the constraints $\Theta_{\CA}^\io $ and $\Theta_{\CA}^\oo$ purely in terms of $\cos \theta_1$, $\cos \theta_2$, and $\cos\phi$:
\begin{align} \label{eq:Thetaclus}
&\Theta_{\CA}^\io (R,\theta_1,\theta_2, \phi) =\,  \theta \Big(\cos \theta_{1} <\cos\theta_{12} \Big) \, \theta\Big( \cos R < \cos\theta_1 < \cos \frac{R}{2} \Big) \,  \nn\\
& \qquad \qquad \qquad \qquad \qquad \quad \! \times \theta\Big( -\cos R < \cos\theta_2 < \cos R \Big) \nn \\
&\Theta_{\CA}^\oo (R,\theta_1,\theta_2, \phi) = \theta\Big( -\cos R <\cos \theta_{1,2} < \cos R \Big)  \, \theta\Big( \cos R<\cos \theta_{12} \Big)  \,,
\end{align}
where the angle $\theta_{12}$ is given by
\begin{align}
\cos\theta_{12} &= \sin \theta_1 \sin \theta_2 \cos \phi+\cos \theta_1 \cos \theta_2 \,,
\end{align}
and we have restricted $\theta_{1}$ and $\theta_{2}$, the angles of $k_1$ and $k_2$ with respect to the $n$-jet direction, to the ranges where clustering can actually take place.  Note that configurations related by the interchange of gluons 1 and 2 or the two jets are accounted for by symmetry factors in $\Delta \ca{M}^{(2)}_\alg$, as in \eq{Mschematic}.

For $\kt$, the constraints on clustering depend on the relative energies of the two gluons.  We note first that the out-out constraint $\Theta_{\kt}^\oo$ is the same as C/A,
\be
\Theta_{\kt}^\oo = \Theta_{\CA}^\oo \,,
\ee
since the constraint only requires that the two gluons are initially outside the jets, are separated by an angle $\theta_{12} < R$, and are clustered into the jets.  The in-out constraint differs, however.  If gluon 1 is in the jet and gluon 2 is out, and they cluster before gluon 1 clusters with the jet, then the constraint is
\be
\min\!\big(E_1^2,E_2^2 \big) (1-\cos\theta_{12}) < E_1^2 (1-\cos\theta_1) \,.
\ee
Clearly if $E_1 < E_2$, then this produces the same constraint as C/A, that $\theta_{12} < \theta_1$.  If $E_2 < E_1$, the constraint on $\theta_{12}$ is now looser:
\be \label{eq:KTclus}
1 - \cos\theta_{12} < \frac{E_1^2}{E_2^2} \, (1-\cos\theta_1) \,.
\ee
This tells us that any time the two gluons will cluster with C/A, they will cluster with $\kt$, and $\kt$ has an additional region of phase space with $\theta_{12} > \theta_1$ and $E_2 < E_1$ where the gluons will cluster.  It is useful to write the in-out constraint for $\kt$, therefore, as
\be \label{eq:ThetaKT}
\Theta_{\kt}^{\io} = \Theta_{\CA}^{\io} + \Theta_{\Delta}^{\io}
\ee
where
\begin{align} \label{eq:ThetaDelta}
\Theta_{\Delta}^{\io} &= \theta \Big(\cos \theta_{12} < \cos \theta_1\Big) \, \theta\Big( z^2 (1-\cos\theta_{12}) < (1-z)^2 (1-\cos\theta_1) \Big) \nn \\
& \qquad \times \theta \Big( \cos R < \cos\theta_1 < 1 \Big) \, \theta\Big( -\cos R < \cos\theta_2 < \cos R \Big) \,.
\end{align}
The constraint that $E_2 < E_1$ is a consequence of the constraints $\theta_{12} > \theta_1$ and the constraint in \eq{KTclus}.  Note that $\Theta_{\Delta}^\io$ is independent of $k_t^0$.

Soft divergences arise from each of the in-out terms $\ca{M}_{\alg}^\io$ at $1/\e^2$ and the out-out terms $\ca{M}_{\alg}^\oo$ at $1/\e$. These divergences come from the integrals over $k^0_1, k^0_2$ (or equivalently $k^0_t,z$).  Whether the combined soft gluon, $k_t$, is in or out of the jet is given by the angle $\theta_{t}$ of $k_t$ with respect to the $n$-jet direction.  If $\theta_{t} < R$, then the clustered gluons are in the jet, and otherwise they are out.  This angle can be written as
\begin{align}\label{eq:cost}
\cos^2 \theta_{t} &= \frac{((1-z) \cos\theta_{1} + z \cos\theta_{2})^2}{1-2 z (1-z) (1- \cos\theta_{12}) } \,.
\end{align}
Note that here $\theta_t$ is independent of $k^0_t$. We can now express the in-out contributions as 
\begin{align} \label{eq:inoutmeas}
\ca{M}^\io_{\rho} &= Q\,\theta\left(\cos^2 R< \cos^2\theta_t\right) \,  \delta\Big(Q\rho - k_t^0 \Big[ (1-z) (1-\cos \theta_1) + z (1-\cos\theta_2) \Big] \Big) \, \delta(\Lambda) \nn \\
\ca{M}^\io_{\Lambda} &= \theta\left(\cos^2 \theta_t < \cos^2 R\right) \,  \delta(\rho) \, \delta \left(\Lambda-k^0_t \right) \nn \\
\ca{M}^\io_{a} &=-  Q\,\delta\Big(Q \rho - k^0_t (1-z) (1-\cos\theta_1) \Big) \, \delta \left(\Lambda - k^0_t z \right) \,,
\end{align}
and the out-out contribution as 
\begin{align}\label{eq:outoutmeas}
 \ca{M}^\oo_{\rho} &=Q\, \theta\left(\cos^2 R< \cos^2\theta_t\right) \, \delta\Big(Q\rho - k_t^0 \Big[ (1-z) (1-\cos \theta_1) + z (1-\cos\theta_2) \Big] \Big) \, \delta( \Lambda)\nn\\
 \ca{M}^\oo_{a} &= -  \theta\left(\cos^2 R< \cos^2\theta_t\right) \,  \delta(\rho) \, \delta \left( \Lambda-k^0_t \right) \,.
\end{align}
Note that these parts of the measurement function are independent of the algorithm, and apply to both C/A and $\kt$.  We find it is easiest to start with the calculation for C/A and then tackle the $\kt$ calculation.

\subsection{Soft Function Correction Term for C/A Dijet Mass Distribution}
\label{ssec:CAmassdist}

We will evaluate the soft function correction for each in-out and out-out measurement contribution in \eq{Mschematic}, where $\Delta S^\io_{i,\,\CA}$ and $\Delta S^\oo_{i,\,\CA}$ correspond to $\ca{M}^\io_i$ and $\ca{M}^\oo_i$ respectively. The $\as^2$ Abelian soft function correction term due to C/A clustering is then expressed as 
\be
\Delta S_{\CA}^{(2)} (\rho,\Lambda) = \left[\Delta S^\io_{\rho,\,\CA}+ \Delta S^\io_{\Lambda,\,\CA}+ \Delta S^\io_{a,\,\CA} \right] + \left[\Delta S^\oo_{\rho,\,\CA}+ \Delta S^\oo_{a,\,\CA} \right] \,.
\ee
Our general approach for each term will be to perform the $k^0_t$ and $z$ integrals to extract the divergences in each term and then to perform the remaining integrals numerically.  We will find that each term can be expanded in terms of finite numeric integrals that have a common form.  This will allow us to show analytically that the divergences in the correction terms cancel and yield a simple form for the result, where we need only numerically calculate coefficients of logs.

\subsubsection{In-out Contribution}
We begin by considering the in-out contributions, where each term $\Delta S^\io_i$ can be written as
\begin{align}\label{eq:DeltaSIO}
\Delta S^\io_{i,\,\CA} &=  \left( \frac{\as C_F}{\pi} \right)^2 \frac{\big( e^{\gamma_E}\big)^{2\e}}{\Gamma(1-\e)^2} \,  \left(\frac{\pi^{1/2} \Gamma(\frac12 - \e)}{\Gamma(1-\e)}\right)^{-1} \int_{-1}^{1} d\cos\phi \,d\cos \theta_1\,d\cos \theta_2  \\
& \qquad \times (1-\cos^2\phi)^{-\frac{1}{2}-\e}(1-\cos^2\theta_1)^{-1-\e}(1-\cos^2\theta_2)^{-1-\e} 8 \, \Theta_{\CA}^\io (R,\theta_1,\theta_2, \phi) \, I_{i}^\io \,,  \nn 
\end{align}
where $I_i^\io$ contains the integral over $k_t^0$ and $z$ from which we shall extract the soft divergences
\be\label{eq:Iio}
I_i^\io= \mu^{4\e} \int_0^\infty dk^0_t \int_0^1 dz\, \left(k_t^0 \right)^{-1-4\e} \, \left( (1-z) z \right)^{-1-2\e}\, \ca{M}^\io_i \,.
\ee
For $\Delta S^\io_{a,\,\CA}$, the anti-$\kt$ term, this is straightforward:
\begin{align}
I_{a}^\io & = \mu^{4\e} \int_0^\infty dk^0_t \int_0^1 dz\, \left(k_t^0 \right)^{-1-4\e} \, \left( (1-z) z \right)^{-1-2\e}\, \ca{M}^\io_{a} \nn\\
& = - \mu^{4\e} Q^{-2\e} \rho^{-1-2\e} \Lambda^{-1-2\e}\, (1-\cos R)^{2\e} \left[ 1 + 2\e \ln \left(\frac{1-\cos\theta_1}{1-\cos R}\right) + \e^2 c_{a}^{(2)} \right] \,.
\end{align}
Here $c_a^{ (2) }$ only contributes to the finite, $\rho$ and $\Lambda$ independent terms in the soft function. 

For the $\Delta S^\io_{\rho,\,\CA}$ and $\Delta S^\io_{\Lambda,\,\CA}$ terms the $z$ integral is more complex since the constraint on $\theta_{tn}$ depends on $z$. Singularities in the $z$ integral exist at $z = 0$ and $z = 1$, corresponding to either gluon becoming soft.  The constraint on $z$ is quadratic if we write it in terms of $\cos^2 \theta_{t}$ and $\cos^2 R$ using \eq{cost}.  Physically, we can see that when $z = 0$, the gluon out of the jet is much softer and we will have $\theta_{t} < R$.  When $z = 1$, the gluon in the jet is much softer and we will have $\theta_{t} > R$.  This means that given $\cos\theta_1, \cos\theta_2$, and $\cos\phi$ there will be some $z_0$ with $0 < z_0 < 1$ that is the crossover point where $\theta_{t} (z = z_0) = R$.  So we can rewrite the constraints on $\theta_{t}$ as
\begin{align}\label{eq:thetat}
\theta( \cos^2 R<\cos^2 \theta_{t}) = \theta\big(0 < z < z_0(\cos\theta_1, \cos\theta_2, \cos\phi) \big) \,, \nn \\
\theta( \cos^2\theta_{t} <\cos^2 R) = \theta\big(z_0(\cos\theta_1, \cos\theta_2, \cos\phi) < z < 1 \big) \,.
\end{align}
We can solve for $z_0$ by solving the quadratic equation, $a z^2 + b z + c = 0$, when $\cos^2 \theta_{tn} = \cos^2 R$, taking the root\footnote{We have argued that a single root exists in the range $z \in (0,1)$.  One can show that this is the smaller root if $a > 0$ and the larger root if $a < 0$.  In either case, this corresponds to choosing the negative sign in the quadratic formula.  The root with the positive sign corresponds to a value of $z$ outside of the physical range.} 
\be
z_0 = \frac{1}{2a}\big( - b - \sqrt{b^2 - 4ac}\big) \,,
\ee
where
\begin{align}\label{eq:abc}
a &= (\cos\theta_{1} - \cos\theta_{2})^2 - 2 \cos^2 R (1- \cos\theta_{12}) \,, \nn \\
b &= -2\cos\theta_{1} (\cos\theta_{1} - \cos\theta_{2}) + 2 \cos^2 R (1-\cos\theta_{12}) \,, \nn \\
c &= \cos^2 \theta_1 - \cos^2 R \,.
\end{align}
Performing the integral over $z$ analytically is still challenging.  Instead, we note that the integral over $k_t^0$ produces a single divergences, and therefore to obtain the divergent and finite terms in the $z$ integral we only need to expand the $z$ integrand over $z$ to $\ca{O}(\e)$.  For the $\Delta S^\io_{\Lambda,\,\CA}$ term, corresponding to the case where the gluons are clustered with C/A and end up outside the jet, the integrals are
\begin{align}\label{eq:IioLambda}
I_\Lambda^\io & =  \mu^{4\e}\int_0^\infty dk^0_t \int_0^1 dz\, \left(k_t^0 \right)^{-1-4\e} \, \left( (1-z) z \right)^{-1-2\e}\, \theta(z_0 < z<1)\, \delta(\rho) \, \delta(\Lambda - k^0_t) \nn \\
& =-\frac{1}{2\e}  \mu^{4\e}\, \Lambda^{-1-4\e} \delta(\rho) \, \bigg[ 1 + 2\e \ln \frac{z_0}{1-z_0} + \e^2 c_{\Lambda}^{(2)} \bigg] \,,
\end{align}
where as with $\Delta S^\io_{a,\,\CA}$, $c^{(2)}_{\Lambda}$ will only contribute to the finite, $\rho$ and $\Lambda$ independent parts of the soft function.  Similarly for $\Delta S^\io_{\rho,\,\CA}$,
\begin{align}
I_{\rho}^\io & = \mu^{4\e}\int_0^\infty dk^0_t \int_0^1 dz\, \left(k_t^0 \right)^{-1-4\e} \, \left( (1-z) z \right)^{-1-2\e}\, \theta(0 < z < z_0) \, \delta(\Lambda) \nn \\
& \quad \times Q \delta\Big(Q\rho - k_t^0 \Big[ (1-z) (1-\cos \theta_1) + z (1-\cos\theta_2) \Big] \Big)  \\
& = -\frac{\mu^{4\e} }{2\e} Q^{-4\e} \delta(\Lambda) \, \rho^{-1-4\e} \, (1-\cos R)^{4\e} \bigg[ 1 + 2\e \bigg( -\ln \frac{z_0}{1-z_0} + 2 \ln \frac{1-\cos\theta_1}{1-\cos R} \bigg) + \e^2 c_{\rho}^{(2)} \bigg] \,. \nn
\end{align}
As above, $c_{\rho}^{(2)}$ will only contribute to the finite, $\rho$ and $\Lambda$ independent parts of the soft function.

Now we have extracted all of the divergences, and only finite integrals remain.  At this point, we assemble the results of the in-out contribution to the soft function, and we will find that we can simplify the result significantly before evaluating the remaining integrals numerically.  For each term $\Delta S^\io_{i,\,\CA}$ expressed as in \eq{DeltaSIO}, after doing the $k_t^0$ and $z$ integrals we have
\begin{align}
I_a^\io &= - \mu^{4\e} Q^{-2\e} \Lambda^{-1-2\e} \rho^{-1-2\e} (1-\cos R)^{2\e} \, \big[ 1 + \e \,c_{ a}^{(1)} + \e^2 c_{ a}^{(2)} \big] \,, \nn \\
I_\Lambda^\io &= -\frac{1}{2\e} \,\mu^{4\e}\Lambda^{-1-4\e} \, \delta(\rho)  \bigg[ 1 + \e \,c_{ \Lambda}^{(1)}  + \e^2 c_{\Lambda}^{(2)} \bigg] \,, \nn \\
I_{\rho}^\io &= -\frac{\mu^{4\e} }{2\e} Q^{-4\e}\rho^{-1-4\e} (1-\cos R)^{4\e} \, \delta(\Lambda)\big[ 1 + \e\, c_{ \rho}^{(1)} + \e^2 c_{ \rho}^{(2)} \big] \,,
\end{align}
where
\begin{align}
c_{a}^{(1)} &= 2\ln \frac{1-\cos\theta_1}{1-\cos R} \,, \nn \\
c_{\Lambda}^{(1)} &= 2\ln \frac{z_0}{1-z_0} \,, \nn \\
c_{\rho}^{(1)} &= -2\ln \frac{z_0}{1-z_0} + 4 \ln \frac{1-\cos\theta_1}{1-\cos R} \,.
\end{align}
We can see that for each term, the integral over the leading order divergence is the same.  At the next order, the $c_{i}^{(1)}$ change the numeric integrals, but they are related since
\be \label{eq:cirelation}
-2c_{ a}^{(1)} + c_{ \Lambda}^{(1)} + c_{ \rho}^{(1)} = 0
\ee
We have added $R$ dependent terms into $c_a^{(1)}$ and $c_{\rho}^{(1)}$ (and compensating terms into the prefactors, which will change the arguments of the NGLs) to ensure that the remaining integrals are finite as $R\to 0$.  We will write the soft function correction in terms of the remaining integrals.  Define
\begin{align} \label{eq:AjcaIO}
A^{(j)}_{i,\,\CA} (R) &= \frac{\big( e^{\gamma_E}\big)^{2\e}}{\Gamma(1-\e)^2} \,  \left(\frac{\pi^{1/2} \Gamma(\frac12 - \e)}{\Gamma(1-\e)}\right)^{-1} \int_{-1}^{1} d\cos\phi \,d\cos \theta_1\,d\cos \theta_2  \\
& \quad \times (1-\cos^2\phi)^{-\frac{1}{2}-\e}(1-\cos^2\theta_1)^{-1-\e}(1-\cos^2\theta_2)^{-1-\e} 8 \, \Theta_{\CA}^\io (R,\theta_1,\theta_2, \phi)  \, c_{ i}^{(j)} \,,\nn
\end{align}
for $j =$ 0, 1, or 2, where $c_ { i} ^{(0)} = 1$ for all $i$ and we just call this integral $A_{\CA}^{(0)} (R)$.  Since we will find the total result is finite, we will be able to set $\e = 0$ in the integrals and perform them numerically.  Using \eq{cirelation}, we have
\be \label{eq:Arelation}
-2A_{a,\,\CA} ^{(1)} + A_{\Lambda,\,\CA} ^{(1)} + A_ {\rho,\,\CA} ^{(1)} = 0
\ee
In terms of these integrals, the in-out contribution to the soft function correction term is very simple:
\begin{align}
\Delta S_{\CA}^\io &= \, \Delta S^\io_{\rho,\,\CA}+  \Delta S^\io_{\Lambda,\,\CA} + \Delta S^\io_{a,\,\CA} \nn \\
&= \left( \frac{\as C_F}{\pi} \right)^2 \frac{1}{Q(1-\cos R)} \nn \\
& \quad \times \bigg\{ 2C_{\CA}^{(0)}(R)  \bigg[ \delta\left(\frac{\rho}{1-\cos R}\right) \ca{L}_1 \left(\frac{\Lambda}{Q} \right) + \delta\!\left(\frac{\Lambda}{Q} \right) \ca{L}_1\left(\frac{\rho}{1-\cos R}\right) \nn \\
& \qquad \qquad \qquad \qquad - \ca{L}_0\left(\frac{\rho}{1-\cos R}\right)\ca{L}_0\!\left( \frac{\Lambda}{Q} \right) \bigg]  \nn \\
& \qquad \quad + C^{(1)}_{\io,\,\CA} (R) \left[ \delta\! \left(\frac{\Lambda}{Q} \right) \ca{L}_0 \left(\frac{\rho}{1-\cos R}\right) - \delta\left(\frac{\rho}{1-\cos R}\right)\ca{L}_0 \left(\frac{\Lambda}{Q} \right) \right] \nn \\
& \qquad \quad + C^{(2)}_{\io,\,\CA} (R) \, \delta\left(\frac{\rho}{1-\cos R}\right)\, \delta\!\left( \frac{\Lambda}{Q} \right)  \bigg\} \,,
\end{align}
where $\ca{L}_0 (x)$ and $\ca{L}_1 (x)$ are distribution functions,
\be
\ca{L}_0 (x) = \left[ \frac{\theta(x)}{x} \right]_+ \,, \quad \ca{L}_1 (x) = \left[ \theta(x)\frac{\ln x}{x} \right]_+ \,.
\ee
The NGL coefficients are
\begin{align}
C_{\CA}^{(0)} (R) &= \frac12 A^{(0)} (R) \,, \\
C_{\io,\,\CA}^{(1)} (R) &= \frac{1}{2}\left(A^{(1)}_{a,\,\CA} (R)-A^{(1)}_ {\rho,\,\CA} (R)\right) \,, \nn \\
C_{\io,\,\CA}^{(2)} (R) &= \frac{1}{8} \left( A^{(2)}_ { \Lambda}(R)+A^{(2)}_ { \rho} (R)-2A^{(2)}_ { a}(R) \right) - \frac{1}{2}\left(A^{(1)}_{ a} (R)-A^{(1)}_ { \rho} (R)\right) \ln (1-\cos R) \,. \nn
\end{align}
We will not explicitly calculate $C_{\io,\,\CA}^{(2)} (R)$.

We can see that the divergences in $\Delta S^\io$ have cancelled, which is a consequence of \eq{Arelation}.  This means we can set $\e=0$ in the $A^{(j)}_{i,\CA}$ integrals.  Note that we have shifted the argument of the single NGL to include a factor of $1 - \cos R$ to give the same argument as the double NGL.  This change simply alters the constant $\rho$ and $\Lambda$ independent terms.

\subsubsection{Out-out Contribution} \label{ssec:OutOut}
Next we consider the out-out contribution to the soft function correction term, represented in \fig{CAps}(c). Each term $\Delta S^\oo_i$ can be written as
\begin{align}\label{eq:DeltaSOO}
\Delta S^\oo_{i\,,\CA} &=  \left( \frac{\as C_F}{\pi} \right)^2 \frac{\big( e^{\gamma_E}\big)^{2\e}}{\Gamma(1-\e)^2} \,  \left(\frac{\pi^{1/2} \Gamma(\frac12 - \e)}{\Gamma(1-\e)}\right)^{-1} \int_{-1}^{1} d\cos\phi \,d\cos \theta_1\,d\cos \theta_2  \\
& \qquad \times (1-\cos^2\phi)^{-\frac{1}{2}-\e}(1-\cos^2\theta_1)^{-1-\e}(1-\cos^2\theta_2)^{-1-\e} 4 \, \Theta_{\CA}^\oo (R,\theta_1,\theta_2, \phi) \, I_i^\oo \,,  \nn 
\end{align}
where as before $I_i^\oo$ contains the integrals over $k_t^0$ and $z$ from which we shall extract the soft divergences:
\be\label{eq:Ioo}
I_i^\oo= \mu^{4\e} \int_0^\infty dk^0_t \int_0^1 dz\, \left(k_t^0 \right)^{-1-4\e} \, \left( (1-z) z \right)^{-1-2\e}\, \ca{M}^\oo_i \,.
\ee
Since both gluons are outside of the cone of radius $R$, as $z\to 0$ or 1 the combined pair $k_t$ will be outside the jet.  Therefore, there are no divergences in the $z$ integral, and the only divergences in the out-out contribution arise when both gluons are soft ($k_t^0 = 0$). As in the in-out case, the constraint $\cos^2 R<\cos^2\theta_t$ can be expressed as a quadratic function of $z$,
\be
\theta( \cos^2 R<\cos^2 \theta_{t}) = \theta\big( z_1(\cos\theta_1, \cos\theta_2, \cos\phi) <z < z_0(\cos\theta_1, \cos\theta_2, \cos\phi) \big) \,, 
\ee    
where
\be
z_{1} = \frac{1}{2a}\big( - b + \sqrt{b^2 - 4ac}\big) \,, \quad z_{0} = \frac{1}{2a}\big( - b - \sqrt{b^2 - 4ac}\big) \,,
\ee
and $a,b,c$ are defined in \eq{abc}. Since $z_0$ and $z_1$ are independent of $k_t^0$ we can carry out this integral and extract the divergent contributions. The contribution from C/A clustering for this configuration is 
\begin{align}
I_{\rho}^\oo & = \mu^{4\e}  \int_0^\infty dk^0_t \int_0^1 dz\, \left(k_t^0 \right)^{-1-4\e} \, \big( (1-z) z \big)^{-1-2\e} \delta(\Lambda)\, \theta(z_1<z<z_0) \nn \\
&\quad \times \delta \! \left( k_t^0-\frac{Q \rho}{ (1-z)(1-\cos\theta_1)+z(1-\cos\theta_2)  } \right)  \big( (1-z)(1-\cos\theta_1)+z(1-\cos\theta_2) \big)^{4\e} \nn \\
& = \frac{1}{Q} \left( \frac{\mu}{Q}\right)^{4\e} \delta \! \left( \frac{\Lambda}{Q} \right) \rho^{-1-4\e} \bigg[ \ln \left( \frac{ (1-z_1) z_0 }{ (1-z_0) z_1 } \right) + \epsilon \, c_{\oo,\,\rho}^{(2)} \bigg]\,,
\end{align}
where we can set the $\e =0$ in the integral over z, since we have argued it is finite. The corresponding terms from anti-$\kt$ for the out-out configuration give
\begin{align}\label{eq:IaOO}
I_a^\oo & = - \mu^{4\e} \int_0^\infty dk^0_t \int_0^1 dz\, \left(k_t^0 \right)^{-1-4\e} \, \big( (1-z) z \big)^{-1-2\e} \delta(\rho)  \, \theta(z_1<z<z_0) \,\delta(\Lambda- k_t^0 )  \nn \\
& = - \frac{1}{Q} \left( \frac{\mu}{Q}\right)^{4\e} \delta(\rho) \left( \frac{\Lambda}{Q} \right)^{-1-4\e} \bigg[ \ln \left( \frac{ (1-z_1) z_0 }{ (1-z_0) z_1 } \right) + \epsilon \, c_{\oo,\,a}^{(2)} \bigg] \,.
\end{align}

We define the remaining angular integrals in the calculation of the out-out soft function to be 
\begin{align}\label{eq:AOOdefn}
A_{\oo,\,i}^{(j)} (R) &= \frac{\big( e^{\gamma_E}\big)^{2\e}}{\Gamma(1-\e)^2} \,  \left(\frac{\pi^{1/2} \Gamma(\frac12 - \e)}{\Gamma(1-\e)}\right)^{-1} \int_{-1}^{1} d\cos\phi \,d\cos \theta_1\,d\cos \theta_2 \, (1-\cos^2\phi)^{-\frac{1}{2}-\e} \nn \\
& \quad \times (1-\cos^2\theta_1)^{-1-\e}(1-\cos^2\theta_2)^{-1-\e} 4 \, \Theta_{\CA}^\oo (R,\theta_1,\theta_2, \phi) \, c_{\oo,\,i}^{(j)}  \,\,,
\end{align}
for $j = 1,2$ and $i = \rho,a$ where
\be
c_{\oo,\,i}^{(1)} = \ln \left( \frac{ (1-z_1) z_0 }{ (1-z_0) z_1 } \right)\,.
\ee
The out-out soft function correction term then takes the simple form,
\begin{align}
\Delta S_{\CA}^\oo &= \, \Delta S^\oo_{\rho,\,\CA}+ \Delta S^\io_{a,\,\CA} \\
&= \left( \frac{\as C_F}{\pi} \right)^2 \frac{1}{Q(1-\cos R)} \, \bigg\{ \nn \\
& \qquad \qquad  C_{\oo,\,\CA}^{(1)} (R) \bigg[ \delta\! \left( \frac{\Lambda}{Q} \right) \ca{L}_0\left(\frac{\rho}{1-\cos R}\right)  - \delta\left(\frac{\rho}{1-\cos R}\right) \ca{L}_0 \left(\frac{\Lambda}{Q} \right) \bigg] \nn \\
& \qquad \qquad + C_{\oo,\,\CA}^{(2)} (R) \, \delta\left(\frac{\rho}{1-\cos R}\right)\delta\! \left( \frac{\Lambda}{Q} \right)\bigg\} \,, \nn
\end{align}
The coefficients are
\begin{align}
C^{(1)}_{\oo,\,\CA} (R) &= A^{(1)}_{\oo} (R) \,, \nn \\
C^{(2)}_{\oo,\,\CA} (R) &= A^{(1)}_{\oo} (R) \ln(1-\cos R) + A^{(2)}_{\oo,\,\rho} (R) - A^{(2)}_{\oo,\,a} (R) \,.
\end{align}
As with the in-out contributions, since the divergences have cancelled we can set $\e=0$ and evaluate the remaining integrals numerically. We do this in the following subsection.  Note that we have shifted the argument of the out-out NGL to equal the argument of the in-out NGLs, which affects only the constant $\rho$ and $\Lambda$ independent finite terms.

\subsubsection{Total Soft Function Correction Term for C/A}
Combining the results from the previous subsections, we find the soft function correction term from clustering is 
\begin{align}
&\Delta S_{\CA}^{(2)}(\rho,\Lambda) = \left( \frac{\as C_F}{\pi} \right)^2 \frac{1}{2Q\sin^2 \frac{R}{2}} \nn \\
& \quad \times \bigg\{ 2C_{\CA}^{(0)}(R) \bigg[ \delta\left(\frac{\rho}{2\sin^2 \frac{R}{2}}\right) \ca{L}_1 \left(\frac{\Lambda}{Q} \right)+ \delta\!\left(\frac{\Lambda}{Q} \right) \ca{L}_1\left(\frac{\rho}{2\sin^2 \frac{R}{2}}\right) - \ca{L}_0\left(\frac{\rho}{2\sin^2 \frac{R}{2}}\right) \ca{L}_0\!\left( \frac{\Lambda}{Q} \right) \bigg]  \nn \\
& \qquad \quad + \left(C_{\io,\,\CA}^{(1)} (R) + C_{\oo,\,\CA}^{(1)} (R)\right)  \bigg[ \delta\! \left(\frac{\Lambda}{Q} \right) \ca{L}_0 \left(\frac{\rho}{2\sin^2 \frac{R}{2}}\right) - \delta\left(\frac{\rho}{2\sin^2 \frac{R}{2}}\right)\ca{L}_0 \left(\frac{\Lambda}{Q} \right) \bigg] \nn \\
& \qquad \quad + \left(C_{\io,\,\CA}^{(2)} (R) + C_{\io,\,\CA}^{(2)} (R) \right) \delta\left(\frac{\rho}{2\sin^2 \frac{R}{2}}\right)\, \delta\!\left( \frac{\Lambda}{Q} \right) \,.
\end{align}
Converting to the singly differential distribution in $\rho$ and the cumulant as follows,
\be
S(\rho) = \int_0^{\Lambda_c} d\Lambda \, S(\rho,\Lambda)  \qquad \qquad   \Sigma_S = \int_0^{\rho_c} d\rho \, S(\rho) \,,
\ee
the $\rho$-dependent terms in the correction to the singly differential distribution and  cumulant are
\begin{align}
\Delta S_{\CA}^{(2)}(\rho) &= \left( \frac{\as C_F}{\pi} \right)^2   \bigg\{ 2C_{\CA}^{(0)}(R) \bigg[ \frac12 \delta(\rho) \ln^2\left(\frac{2\Lambda_c}{Q} \sin^2 \frac{R}{2} \right)+ \ca{L}_1(\rho)-\ca{L}_0(\rho)\ln\left( \frac{2\Lambda_c}{Q} \sin^2 \frac{R}{2} \right) \bigg]  \nn \\
& \qquad \qquad \qquad \quad + \left(C_{\io,\,\CA}^{(1)} (R) + C_{\oo,\,\CA}^{(1)} (R)\right) \bigg[ \ca{L}_0 (\rho) - \delta(\rho)\ln\left(\frac{2\Lambda_c}{Q} \sin^2 \frac{R}{2} \right) \bigg] \nn \\
& \qquad \qquad \qquad \quad + \left(C_{\io,\,\CA}^{(2)} (R) + C_{\io,\,\CA}^{(2)} (R) \right) \delta(\rho)\, \,,
\end{align}
\begin{align}
\Delta \Sigma_{S,\,\CA}^{(2)} &= \left( \frac{\as C_F}{\pi} \right)^2 \bigg\{ C_{\CA}^{(0)} (R) \ln^2 \frac{Q\rho_c}{2\Lambda_c \sin^2 \frac{R}{2}} + \left(C_{\io,\,\CA}^{(1)} (R) + C_{\oo,\,\CA}^{(1)} (R)\right) \ln \frac{Q\rho_c}{2\Lambda_c \sin^2 \frac{R}{2}} \nn \\
& \qquad \qquad \qquad \qquad + \text{ constant} \bigg\} \,,
\end{align}
where the constant are two loop finite terms that we do not determine.  Since the total result is finite, we can set $\e = 0$ in the $A^{(0)}, A^{(1)}$ and $A_\oo$ integrals and perform them numerically.  These integrals reduce to
\begin{align}\label{eq:finiteAs}
A^{(j)}_{i,\,\CA} (R) &= \frac{8}{\pi} \int_{-1}^{1} d\cos\phi \,d\cos \theta_1\,d\cos \theta_2 \, \frac{  (1-\cos^2\phi)^{-\frac{1}{2}} }{(1-\cos^2\theta_1) (1-\cos^2\theta_2)}  \Theta_{\CA}^\io (R,\theta_1,\theta_2, \phi)  \, c_{i}^{(j)} \nn \\
A_{\oo}(R) &= \frac{4}{\pi} \int_{-1}^{1} d\cos\phi \,d\cos \theta_1\,d\cos \theta_2 \, \frac{  (1-\cos^2\phi)^{-\frac{1}{2}} }{(1-\cos^2\theta_1) (1-\cos^2\theta_2)}  \Theta_{\CA}^\oo (R,\theta_1,\theta_2, \phi)  \, c_\oo \,.
\end{align}
These determine the coefficients of the clustering NGLs, which we plot in the main text in \fig{NGLcoeffs}.

\section{Soft Function Correction Term for the $\kt$ Dijet Mass Distribution}

The soft function correction term for $\kt$ also comes in two parts, an in-out contribution and an out-out contribution.  As discussed previously, the out-out contribution is the same as C/A, and was calculated above.  The in-out contribution for $\kt$ differs, and can be written as the in-out contribution from C/A plus an additional term, as in \eq{ThetaKT}.  The clustering constraints for this additional term, $\Theta_{\Delta}^{\io}$, are defined in \eq{ThetaDelta}.  We will calculation the soft function correction term, $\Delta S_{\Delta}^{\io}$, that corresponds to $\Theta_{\Delta}^{\io}$.  Just as for C/A, this has the form
\be \label{eq:SDelta}
\Delta S^{\io}_{\Delta} = \Delta S^{\io}_{\rho,\,\Delta} + \Delta S^{\io}_{\Lambda,\,\Delta} + \Delta S^{\io}_{a,\,\Delta} \,,
\ee
with
\begin{align}\label{eq:DeltaSDelta}
\Delta S^\io_{i,\,\Delta} &=  \left( \frac{\as C_F}{\pi} \right)^2 \frac{\big( e^{\gamma_E}\big)^{2\e}}{\Gamma(1-\e)^2} \,  \left(\frac{\pi^{1/2} \Gamma(\frac12 - \e)}{\Gamma(1-\e)}\right)^{-1} \int_{-1}^{1} d\cos\phi \,d\cos \theta_1\,d\cos \theta_2  \\
& \qquad \times (1-\cos^2\phi)^{-\frac{1}{2}-\e}(1-\cos^2\theta_1)^{-1-\e}(1-\cos^2\theta_2)^{-1-\e} 8 \, \Theta_{\text{PS},\Delta}^\io (R,\theta_1,\theta_2, \phi) \, I_{i,\,\Delta}^\io \,,  \nn 
\end{align}
where $\Theta_{\text{PS},\Delta}^\io$ is equal to $\Theta_{\Delta}^\io$ with the $z$-dependent $\kt$ clustering constraint removed,
\be
\Theta_{\text{PS},\Delta}^\io = \theta \Big(\cos \theta_{12} < \cos \theta_1\Big) \, \theta \Big( \cos R < \cos\theta_1 < 1 \Big) \, \theta\Big( -\cos R < \cos\theta_2 < \cos R \Big) \,,
\ee
and $I_{i,\,\Delta}^\io$ contains the integrals over $k_t^0$ and $z$ from which we shall extract the soft divergences,
\begin{align}
I_{i,\,\Delta}^\io &= \mu^{4\e} \int_0^\infty dk^0_t \int_0^1 dz\, \left(k_t^0 \right)^{-1-4\e} \, \left( (1-z) z \right)^{-1-2\e} \nn \\
& \qquad \qquad \qquad \qquad \times \theta\Big(z^2 (1 - \cos\theta_{12}) < (1-z)^2 (1-\cos\theta_1) \Big) \,\ca{M}^\io_i \,.
\end{align}
The clustering constraint that we extracted from $\Theta_{\Delta}^{\io}$ and put into $I_{i,\,\Delta}^\io$ can be written as a constraint on $z$:
\be
\theta\Big(z^2 (1 - \cos\theta_{12}) < (1-z)^2 (1-\cos\theta_1) \Big) = \theta( z < z_{\kt} ) \,,
\ee
where
\be
z_{\kt} = \left( 1 + \sqrt{\frac{1 - \cos\theta_{12}}{1-\cos\theta_1}} \right)^{-1} \,.
\ee
This means that the calculation of $I_{i,\,\Delta}^{\io}$ terms is the same as the $I_{i}^{\io}$ terms for C/A with the constraint $z < z_{\kt}$.  This is a simple change to these calculations, and the results are
\begin{align}
I_{a,\,\Delta}^\io & = - \mu^{4\e} Q^{-2\e} \rho^{-1-2\e} \Lambda^{-1-2\e}\, (1-\cos R)^{2\e} \, \theta\left( \frac{\sin\frac{\theta_1}{2} \sin \frac{\theta_{12}}{2}}{\sin^2 \frac{R}{2}} < \frac{Q\rho}{2\Lambda \sin^2 \frac{R}{2}} \right) \nn \\
& \qquad \qquad \qquad \times \left[ 1 + 2\e \ln \left(\frac{1-\cos\theta_1}{1-\cos R}\right) + \e^2 c_{a,\,\Delta}^{(2)} \right] \,,
\end{align}
\begin{align}
I_{\rho,\,\Delta}^{\io} &= -\frac{\mu^{4\e} }{2\e} Q^{-4\e} \delta(\Lambda) \, \rho^{-1-4\e} \, (1-\cos R)^{4\e} \nn \\
& \qquad \qquad \times \bigg[ 1 + 2\e \bigg( -\ln \frac{z_{\rho}}{1-z_{\rho}} + 2 \ln \frac{1-\cos\theta_1}{1-\cos R} \bigg) + \e^2 c_{\rho,\,\Delta}^{(2)} \bigg] \,,
\end{align}
where
\be
z_{\rho} = \min(z_0, z_{\kt}) \,,
\ee
and
\begin{align}
I_{\Lambda,\,\Delta}^\io &= \mu^{4\e}\, \Lambda^{-1-4\e} \delta(\rho) \, \bigg[ \ln \frac{(1-z_0)z_{\kt}}{(1-z_{\kt})z_0} + \e \, c_{\delta,\,\Delta}^{(2)} \bigg] \theta(z_0 < z_{\kt}) \,.
\end{align}
Note that $I_{a,\,\Delta}^{\io}$ has $\rho$ and $\Lambda$ dependence that is not simply a distribution in each variable, since there is a constraint on the relative values.  This means it is difficult to expand the contributions out in terms of distribution (plus) functions.  Therefore we will only determine the double cumulant soft function for the $\kt$ algorithm, which is much simpler.  We are free to move to the cumulant before performing the remaining finite integrals, and for each contribution we obtain
\begin{align}
I_{a,\,\Delta}^{\io,\,\Sigma} &= -\left(\frac{\mu}{Q}\right)^{4\e} \frac{1}{8\e^2} \bigg\{ \theta\left( \frac{\sin\frac{\theta_1}{2} \sin \frac{\theta_{12}}{2}}{\sin^2 \frac{R}{2}} < \frac{Q\rho_c}{2\Lambda_c \sin^2 \frac{R}{2}} \right) \nn \\
& \qquad \quad \times \left(\frac{\Lambda_c}{Q}\right)^{-2\e} \bigg[ 2 \left(\frac{\rho_c}{1-\cos R}\right)^{-2\e} - \left(\frac{\Lambda_c}{Q}\right)^{-2\e} \left( \frac{\sin \frac{\theta_1}{2} \sin \frac{\theta_{12}}{2}}{\sin^2 \frac{R}{2}}\right)^{-2\e} \bigg] \nn \\
& \qquad \qquad + \theta\left( \frac{\sin\frac{\theta_1}{2} \sin \frac{\theta_{12}}{2}}{\sin^2 \frac{R}{2}} > \frac{Q\rho_c}{2\Lambda_c \sin^2 \frac{R}{2}} \right) \left(\frac{\rho_c}{1-\cos R}\right)^{-2\e} (1-\cos R)^{-2\e} \bigg\} \nn \\
& \qquad \times \left[ 1 + 2\e \ln \left(\frac{1-\cos\theta_1}{1-\cos R}\right) + \e^2 c_{a,\,\Delta}^{(2)} \right] \,,
\end{align}
\begin{align}
I_{\rho,\,\Delta}^{\io,\,\Sigma} &= \left(\frac{\mu}{Q}\right)^{4\e} \frac{1}{8\e^2} \left( \frac{\rho_c}{1-\cos R}\right)^{-4\e} \nn \\
& \qquad \qquad \times \bigg[ 1 + 2\e \bigg( -\ln \frac{z_{\rho}}{1-z_{\rho}} + 2 \ln \frac{1-\cos\theta_1}{1-\cos R} \bigg) + \e^2 c_{\rho,\,\Delta}^{(2)} \bigg] \,,
\end{align}
\begin{align}
I_{\Lambda,\,\Delta}^{\io,\,\Sigma} &= -\left(\frac{\mu}{Q}\right)^{4\e} \frac{1}{4\e} \left(\frac{\Lambda_c}{Q} \right)^{-4\e} \bigg[ \ln \frac{(1-z_0)z_{\kt}}{(1-z_{\kt})z_0} + \e \, c_{\Lambda,\,\Delta}^{(2)} \bigg] \theta(z_0 < z_{\kt}) \,.
\end{align}
If we add the cumulants, then we find
\begin{align}
& I_{a,\,\Delta}^{\io,\,\Sigma} + I_{\rho,\,\Delta}^{\io,\,\Sigma} + I_{\Lambda,\,\Delta}^{\io,\,\Sigma} \nn \\
& \qquad = \theta\left( \frac{\sin\frac{\theta_1}{2} \sin \frac{\theta_{12}}{2}}{\sin^2 \frac{R}{2}} < \frac{Q\rho_c}{2\Lambda_c \sin^2 \frac{R}{2}} \right) \nn \\
& \qquad \qquad \times \bigg[ \frac12 \ln^2 \left( \frac{Q\rho_c}{2\Lambda_c \sin^2 \frac{R}{2}}\right) + \ln \left( \frac{Q\rho_c}{2\Lambda_c \sin^2 \frac{R}{2}}\right) \ln \left( \frac{z_{\kt}}{1-z_{\kt}} \frac{1-\cos\theta_1}{1-\cos R} \right) \bigg] \nn \\
& \qquad \quad + \ln \left( \frac{Q\rho_c}{2\Lambda_c \sin^2 \frac{R}{2}}\right) \ln\left(\frac{z_0 (1 - z_{\kt})}{z_{\kt} ( 1-z_0)} \right) \theta(z_0 < z_{\kt}) + \, \text{constant.}
\end{align}
Note that all the divergences cancel.  The constant depends only on $R$, and we do not calculate it.  Therefore, as with C/A, we can set $\e = 0$ and define the coefficients in terms of finite integrals which we will evaluate numerically.

The coefficients are related to the integrals
\begin{align} \label{eq:ADelta}
A_{\Delta}^{(j)} &=  \int_{-1}^{1} d\cos\phi \,d\cos \theta_1\,d\cos \theta_2 (1-\cos^2\phi)^{-\frac{1}{2}}(1-\cos^2\theta_1)^{-1}(1-\cos^2\theta_2)^{-1} \nn \\
& \qquad \times \frac{8}{\pi} \, \Theta_{\text{PS},\Delta}^\io (R,\theta_1,\theta_2, \phi) \, c_{\Delta}^{(j)} \,,
\end{align}
where
\begin{align} \label{eq:cDelta}
c_{\Delta}^{(0)} &= \frac12 \, \theta\left( \frac{\sin\frac{\theta_1}{2} \sin \frac{\theta_{12}}{2}}{\sin^2 \frac{R}{2}} < \frac{Q\rho_c}{2\Lambda_c \sin^2 \frac{R}{2}} \right) \,, \\
c_{\Delta}^{(1)} &= \theta\left( \frac{\sin\frac{\theta_1}{2} \sin \frac{\theta_{12}}{2}}{\sin^2 \frac{R}{2}} < \frac{Q\rho_c}{2\Lambda_c \sin^2 \frac{R}{2}} \right) \ln \left( \frac{z_{\kt}}{1-z_{\kt}} \frac{1-\cos\theta_1}{1-\cos R} \right) \nn \\
& \qquad + \ln\left(\frac{z_0 (1 - z_{\kt})}{z_{\kt} ( 1-z_0)} \right) \theta(z_0 < z_{\kt}) \,. \nn
\end{align}
Then the soft function cumulant $\Delta \Sigma^{\io}_{S,\,\Delta}$ (which is the cumulant of $S^{\io}_{\Delta}$ defined in \eq{SDelta}) is
\be
\Delta S^{\io}_{\Delta} = A_{\Delta}^{(0)} \ln^2 \left( \frac{Q\rho_c}{2\Lambda_c \sin^2 \frac{R}{2}}\right) + A_{\Delta}^{(1)} \ln \left( \frac{Q\rho_c}{2\Lambda_c \sin^2 \frac{R}{2}}\right) + \text{ constant.}
\ee
Finally, we can add these terms to the soft function cumulant for C/A and obtain the result for $\kt$:
\begin{align}
\Delta \Sigma_{S,\,\kt}^{(2)} &= \left( \frac{\as C_F}{\pi} \right)^2 \bigg\{ C_{\kt}^{(0)} (R) \ln^2 \frac{Q\rho_c}{2\Lambda_c \sin^2 \frac{R}{2}} + \left(C_{\io,\,\kt}^{(1)} (R) + C_{\oo,\,\kt}^{(1)} (R)\right) \ln \frac{Q\rho_c}{2\Lambda_c \sin^2 \frac{R}{2}} \nn \\
& \qquad \qquad \qquad \qquad + \text{ constant} \bigg\} \,,
\end{align}
where
\begin{align}
C_{\kt}^{(0)} &= C_{\CA}^{(0)} + A_{\Delta}^{(0)} \,, \nn \\
C_{\io,\,\kt}^{(1)} &= C_{\io,\,\CA}^{(1)} + A_{\Delta}^{(1)} \,, \nn \\
C_{\oo,\,\kt}^{(1)} &= C_{\oo,\,\CA}^{(1)} \,.
\end{align}
Note that the in-out $\kt$ coefficients implicitly depend on the values of $\rho_c$ and $\Lambda_c$ through $A_{\Delta}^{(0,1)}$.  In particular, they depend on the value of the argument of the NGL, $Q\rho_c / 2\Lambda_c \sin^2 \frac{R}{2}$.  From \eq{cDelta}, we can see that when this quantity is larger than 1, the $\rho_c$ and $\Lambda_c$ dependent constraint will always be satisfied, since $\theta_1 < R$ and $\theta_{12} < R$.  As the NGL argument decreases, it will shrink the phase space over which the coefficients contribute, reducing their value.  These coefficients are plotted as a function of $R$ in \fig{NGLcoeffs} for the range $Q\rho_c / 2\Lambda_c \sin^2 \frac{R}{2} \ge 1$.

\bibliography{../../jets}

\end{document}